\documentclass[aps,
prd,
twocolumn,
superscriptaddress,
showpacs,
preprintnumbers,
nofootinbib,
floatfix]{revtex4-1}

\usepackage[export]{adjustbox}
\usepackage[english]{babel}
\usepackage[utf8]{inputenc}
\usepackage{color,graphicx}
\usepackage{bm,bbm}
\usepackage{enumerate}
\usepackage{hyperref}
\usepackage{mathtools}
\usepackage{amssymb}
\usepackage{mathrsfs}
\usepackage{mciteplus}
\usepackage{slashed}
\usepackage{soul}
\usepackage{xspace}
\usepackage{orcidlink}
\usepackage{relsize}
\usepackage{amsmath}
\usepackage{amsthm}
\usepackage{stfloats}
\usepackage{lipsum}

\newcommand{\diff}[0]{\mathrm{d}}

\renewcommand{\k}{\bm{k}}
\newcommand{\p}{\bm{p}}
\newcommand{\q}{\bm{q}}

\renewcommand{\P}{\bm{P}}

\newcommand{\n}{\bm{n}}

\newcommand{\one}{\mathbbm{1}}

\newcommand{\Dc}{\mathcal{D}}

\newcommand{\Kc}{\mathcal{K}}
\newcommand{\Lc}{\mathcal{L}}
\newcommand{\Mc}{\mathcal{M}}
\newcommand{\Nc}{\mathcal{N}}
\newcommand{\Oc}{\mathcal{O}}

\newcommand{\Rc}{\mathcal{R}}

\newcommand{\Tc}{\mathcal{T}}

\newcommand{\Yc}{\mathcal{Y}}

\DeclareMathOperator{\im}{Im}

\newcommand{\bcol}{\left[ \begin{array}{c}}
\newcommand{\ecol}{\end{array} \right]}
\newcommand{\beq}{\begin{eqnarray}}
\newcommand{\eeq}{\end{eqnarray}}

\newcommand{\kev}{\ensuremath{{\mathrm{\,ke\kern -0.1em V}}}\xspace}
\newcommand{\mev}{\ensuremath{{\mathrm{\,Me\kern -0.1em V}}}\xspace}
\newcommand{\gev}{\ensuremath{{\mathrm{\,Ge\kern -0.1em V}}}\xspace}
\newcommand{\tev}{\ensuremath{{\mathrm{\,Te\kern -0.1em V}}}\xspace}

\usepackage{mfirstuc} 
\newcommand{\addReviewer}[2]{
  \expandafter\newcommand\csname #1\endcsname[1]{{\bf \color{#2} \capitalisewords{#1}:\,##1}}
  \expandafter\newcommand\csname #1cor\endcsname[2]{{\color{#2} \capitalisewords{#1}:\,\st{##1}{\bf ##2}}}
  \expandafter\newcommand\csname #1color\endcsname{#2}
}

\usepackage{soul,color}
\definecolor{chromeyellow}{rgb}{1.0, 0.65, 0.0}
\definecolor{DodgeBlue}{rgb}{0.118, 0.565,1.000}
\definecolor{asparagus}{rgb}{0.53, 0.66, 0.42}
\definecolor{cardinal}{rgb}{0.77, 0.12, 0.23}
\definecolor{cadmiumgreen}{rgb}{0.0, 0.42, 0.24}
\definecolor{applegreen}{rgb}{0.55, 0.71, 0.0}

\addReviewer{wrong}{red}
\addReviewer{good}{green}
\addReviewer{sebastian}{cardinal}
\addReviewer{andrew}{DodgeBlue}

\hypersetup{ 
pdfnewwindow=true,
colorlinks=true,    
linkcolor=blue,
citecolor=blue,
filecolor=blue,
urlcolor=blue
} 
 
\begin{document}

\preprint{JLAB-THY-24-4216}

\newcommand{\wm}{Department of Physics, William \& Mary, Williamsburg, VA 23187, USA}

\newcommand{\uw}{Department of Physics, University of Washington, WA 98195, USA}

\newcommand{\ceem}{Center for  Exploration  of  Energy  and  Matter,  Indiana  University,  Bloomington,  IN  47403,  USA}

\newcommand{\indiana}{Physics  Department,  Indiana  University,  
Bloomington,  IN  47405,  USA}

\newcommand{\jlab}{Theory Center, Thomas  Jefferson  National  Accelerator  Facility,  Newport  News,  VA  23606,  USA}


\title{Finite-volume quantization condition from the $N/D$ representation}


\author{Sebastian~M.~Dawid\orcidlink{0000-0001-8498-5254}}
\email[email: ]{dawids@uw.edu}
\affiliation{\uw}

\author{Andrew~W.~Jackura\orcidlink{0000-0002-3249-5410}}
\email[email: ]{awjackura@wm.edu}
\affiliation{\wm}

\author{Adam~P.~Szczepaniak\orcidlink{0000-0002-4156-5492}}
\email[email: ]{aszczepa@indiana.edu}
\affiliation{\jlab}
\affiliation{\indiana}
\affiliation{\ceem}


\begin{abstract}
We propose a new model-independent method for determining hadronic resonances from lattice QCD. The formalism is derived from the general principles of unitarity and analyticity, as encoded in the $N/D$ representation of a partial-wave two-body amplitude. The associated quantization condition relates the finite-volume spectrum to the infinite-volume numerator, $\Nc$, used to reconstruct the scattering amplitude from dispersive relations. Unlike the original L\"uscher condition, this new formalism is valid for energies coinciding with the left-hand cuts from arbitrary one- and multi-particle exchanges.
\end{abstract}

\date{\today}
\maketitle


\emph{Introduction:} The study of hadronic resonances, including exotic states, is crucial for understanding the strong interaction and its effects in many aspects of nuclear and particle physics. Examples include the $\sigma$ meson~\cite{Pelaez:2015qba, Pelaez:2021dak}, the spin-exotic $\pi_1$ hybrid candidate~\cite{IHEP-Brussels-LosAlamos-AnnecyLAPP:1988iqi, JPAC:2018zyd, Woss:2020ayi}, and heavy tetraquark and pentaquark candidates like the $Z_c$~\cite{BESIII:2013ris, Belle:2013yex, Maiani:2013nmn, Pilloni:2016obd} and the $P_c$~\cite{LHCb:2015yax, HillerBlin:2016odx, LHCb:2019kea, Fernandez-Ramirez:2019koa}. These states not only challenge our understanding of how hadrons emerge from quark and gluon interactions in Quantum Chromodynamics (QCD), but impact other fields such as searching for physics beyond the Standard Model by probing the muonic $g-2$~\cite{RBC:2018dos, Borsanyi:2020mff, Muong-2:2021ojo, Muong-2:2023cdq, ExtendedTwistedMass:2022jpw, Athron:2022qpo, Gerardin:2023naa, Davier:2023cyp} or signatures of CP violation~\cite{LHCb:2014mir, AlvarengaNogueira:2015wpj, LHCb:2019jta}.

Lattice QCD has become a powerful tool for rigorously investigating resonances. Although it can only directly access the spectrum of QCD states bound in the finite, discretized Euclidean volume, the work of L\"uscher and others~\cite{Luscher:1986n2, Luscher:1991n1, Rummukainen:1995vs, Kim:2005gf, He:2005ey} on the so-called finite-volume (FV) quantization conditions (QCs) established a framework for relating these energies to infinite-volume (IV) scattering information. Due to these relations, one can quantitatively constrain partial-wave amplitudes and, by analytically continuing them to the complex energy plane, determine the location of resonance poles. In recent years, this formalism has been significantly advanced for applications involving reactions of two- and three-particles~\cite{Davoudi:2011md, Briceno:2012yi, Briceno:2013lba, Briceno:2014oea, Briceno:2012rv, Romero-Lopez:2018zyy, Polejaeva:2012ut, Hansen:2014eka, Hansen:2015zga, Briceno:2017tce, Hammer:2017kms, Mai:2017bge, Briceno:2018aml, Blanton:2020gmf, Hansen:2020zhy, Meng:2021uhz, Muller:2021uur, Blanton:2021mih, Jackura:2022gib} as well as few-hadron matrix elements~\cite{Lellouch:2000pv, Christ:2005gi, Hansen:2012tf, Briceno:2014uqa, Briceno:2015csa, Bernard:2012bi, Briceno:2012yi, Briceno:2015tza, Baroni:2018iau, Briceno:2019nns, Briceno:2019opb, Briceno:2020vgp, Briceno:2020xxs, Briceno:2021xlc, Sherman:2022tco, Lozano:2022kfz} and applied to many important scattering reactions; see~\cite{Briceno:2017max, Hansen:2019nir, Davoudi:2020ngi, Mai:2021lwb} for reviews.

\begin{figure}[h!]
    \centering
    \includegraphics[width=0.48\textwidth]{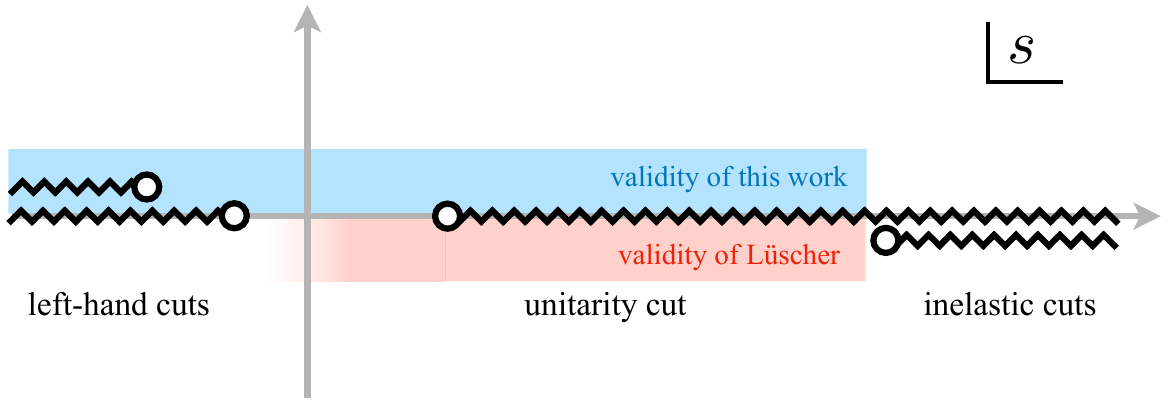}
    \caption{Schematic representation of the partial-wave amplitude branch cut singularities in the complex $s$ plane. The original L\"uscher condition is valid in the region above the nearest left-hand singularity (shown in red).  
    This work extends it to arbitrarily low energies (shown in blue).
    }
    \label{fig:complex_s_plane}
\end{figure}

Despite its successes, L\"uscher’s QC has limitations; it does not account for the effects of virtual particle exchanges in crossed channels. Formally, these appear as branch cuts of amplitudes at energies below the physical threshold and are referred to as the left-hand singularities. Thus, the QC is not applicable at energies that fall below that corresponding to the lightest particle exchange---a region often probed in lattice QCD simulations~\cite{Green:2021qol, Green:2022rjj, Padmanath:2022cvl, Whyte:2024ihh}. Several authors proposed various methods for incorporating one-particle exchange effects in the FV computations~\cite{Klos:2016fdb, Meng:2023bmz, Raposo:2023oru, Bubna:2024izx, Hansen:2024ffk}. In this work, we present a new FV formalism that may include arbitrary one- and multi-particle left-hand cuts, simplifying and generalizing the available approaches; see Fig.~\ref{fig:complex_s_plane}.

We base our approach on the $N/D$ representation of the scattering amplitude~\cite{Omnes:1958hv, Chew:1960iv, Bjorken:1960zz}. It is a general parametrization based on dispersion relations, satisfying the requirements of the S-matrix unitarity and analyticity. In this representation, the partial-wave amplitude $\Mc$ is written as a fraction, $\Nc/\Dc$, where the numerator, $\Nc$, is defined to contain the left-hand singularities, and the denominator, $\Dc$, contains the right-hand unitarity cuts. The $N/D$ representation has been applied extensively in experimental analyses to parametrize scattering data in a way that respects fundamental S-matrix principles~\cite{Badalian:1981xj, Oller:1998zr, Albaladejo:2012sa, JPAC:2017dbi, JPAC:2018zyd}. Here, we show how to employ it in the context of FV lattice computations. We derive a quantization condition that relates the FV energy spectrum, given by the zeros of the FV denominator function, $\Dc_L$, to the FV numerator, $\Nc_L$. We demonstrate that $\Nc_L$ matches the IV representation, $\Nc$, up to corrections exponentially suppressed with the volume, which allows us to replace $\Nc_L \to \Nc$ in the quantization condition. Thus, provided a suitable model of the left-hand cuts is available, the quantization condition allows one to constrain $\Nc$ and then reconstruct the full scattering amplitude from the well-known $N/D$ integral equations. For simplicity, we derive the formalism for a single-channel $2 \to 2$ reaction of spinless particles. The derived condition is applicable at energies arbitrarily below the threshold and, thus, extends the applicability of the canonical FV formalism.

The structure of the paper is as follows. We begin by reviewing the analytic properties of the scattering amplitudes in the $N/D$ representation. This is followed by the derivation of the new quantization condition. We conclude with an overview of the method and discuss its potential extensions. Some technical details are relegated to the supplemental material.


\emph{Infinite-volume scattering:} To illustrate our idea, let us first review the $N/D$ parametrization in the infinite volume. We consider a single-channel elastic scattering of two scalar particles of masses $m_i$, where $i \in \{1,2\}$. They collide with initial four-momenta $p_i = (\omega_i(\p), \p_i)$ and emerge from the reaction with final four-momenta $p'_i = (\omega_i(\p'), \p_i')$. Here, $\omega_i(\p) = (\p_i^2 + m_i^2)^{1/2}$ is the single-particle on-shell energy. The total four-momentum $P = (E, \P) = p_1 + p_2$ and the invariant mass squared $s = P^2$. Other Mandelstam variables are $t = (p_1 - p'_1)^2$ and $u = (p_1 - p'_2)^2$, and they satisfy the relation $s + t + u = 2(m_1^2 + m_2^2)$. In the center-of-momentum (c.m.)~frame (denoted by a ``$\star$'' superscript) the total four-momentum is $P^\star = (E^\star, \bm 0)$ and initial (final) particle has momentum $\p^\star \equiv \p_1^\star = -\p_2^\star$ (${\p'}^\star \equiv {\p'_1}^\star = -{\p_2'}^\star$). Its magnitude is given by,
    \begin{align}
    \label{eq:momentum}
    p^\star = |\p^\star| = |{\p'}^\star| = \frac{1}{2\sqrt{s}} \, \lambda^{1/2}(s, m_1^2, m_2^2) \, ,
    \end{align}
where $\lambda(x,y,z) = x^2 + y^2 + z^2 - 2xy - 2 yz - 2zx$ is the K\"all\'en triangle function. The total scattering energy is given by $E = \sqrt{ (E^\star)^2 + \P^2}$ and energy conservation implies that $E^\star = \omega_1(\p^\star) + \omega_{2}(\p^\star)$.

Given the S matrix, $S = \one + i T$, the scattering amplitude is an element of the transition operator,
    \begin{align}
    \langle \p_1' \p_2' | \, T \, | \p_1 \p_2 \rangle = (2\pi)^4 \, \delta^{(4)}(P' - P) \, \Mc(\p', \p; \P) \, .
    \end{align}
To simplify comparison with the FV formalism, we write the amplitude as a function of nine variables, $\p'$, $\p$, $\P$, many of which are redundant. Given energy-momentum conservation and Lorentz invariance of the amplitude in the infinite volume, one can reduce this set of degrees of freedom to two variables, e.g., $s$ and $t$, or $p^\star$ and $ \hat{\p}'^\star \cdot \hat{\p}^\star =\cos \theta^\star$, where $\theta^\star$ is the scattering angle in the c.m.~frame, $\Mc(\p', \p; \P) \equiv \Mc(p^\star, \hat{\p}'^\star \cdot \hat{\p}^\star )$.

Probability conservation enforces that transition operators satisfy the unitarity relation $T - T^\dag = iT^\dag T$. For scattering amplitudes, this gives the unitarity condition,
\begin{widetext}
    \begin{align}
    2 \im \Mc(\p',\p; \P) = \xi \int\! \frac{\diff \k}{(2\pi)^3} \, \frac{\pi \delta\big(E - \omega_1(\k)  - \omega_2(\P-\k) \big) }{2\omega_1(\k) \, 2\omega_2(\P-\k)} \, 
    \Mc^*(\k,\p';\P)  \, \Mc(\k,\p; \P) \, ,
    \label{eq:IV-unitarity}
    \end{align}
\end{widetext}
where we normalize the single-particle scattering states according to $\langle \p'_i | \p_j \rangle =  (2\pi)^3 \,  2 \omega_i(\p) \, \delta_{ij} \, \delta^{(3)}(\p - \p')$. The symmetry factor $\xi$ is $1/2$ if the particles are identical and $1$ otherwise. We also restrict the energy domain to the elastic scattering region, $E^\star \le E^\star_{\mathrm{inel.}}$, where $E^\star_{\mathrm{inel.}}$ is the first inelastic threshold, see Fig.~\ref{fig:complex_s_plane}.

The scattering amplitude is decomposed into partial waves,
\begin{widetext}
    \begin{align}
    \Mc(\p',\p; \P) = 4 \pi \, 
    \sum_{\ell' m_\ell'} \sum_{\ell m_\ell}
    Y^*_{\ell' m_\ell'}(\hat{\p}'^\star) \, 
    \Mc_{\ell' m_\ell'; \ell m_\ell}(s) \, 
    Y_{\ell m_\ell}(\hat{\p}^\star) 
    = \sum_{\ell} (2\ell + 1) \, \Mc_{\ell}(s) \, P_\ell(\hat{\p}'^\star \cdot \hat{\p}^\star) \, ,
    \label{eq:IV-PW}
    \end{align}
\end{widetext}
where angular momentum $\ell \in \mathbb{N}$ and its $z$-axis projection $ -\ell \leq m_\ell \leq \ell$. In the first equality, we used Lorentz invariance to write the partial-wave projection as a function of $s$, and in the second equality we used angular momentum conservation to write $\Mc_{\ell' m_\ell'; \ell m_\ell}(s) = \delta_{\ell' \ell} \, \delta_{m_\ell' m_\ell} \, \Mc_\ell(s)$ as well as the addition theorem for spherical harmonics. It is convenient to extract from the amplitude factors representing threshold suppression at high angular momenta; therefore we define $\Mc_\ell(s) \equiv (p^\star)^{2\ell} \, \tilde{\Mc}_\ell(s)$. Introducing this expansion in Eq.~\eqref{eq:IV-unitarity}, transforming the integral to the c.m.~frame, and evaluating the integral yields the partial-wave unitarity relation,
    \begin{align}
    \im \tilde{\Mc}_\ell(s) = \rho_\ell(s) \, \lvert\tilde{\Mc}_\ell(s)\rvert^2 \, \Theta(s - s_{\mathrm{thr}}) \, ,
    \label{eq:IV-PW-unitarity}
    \end{align}
where $s_{\mathrm{thr}} = (m_1 + m_2)^2$ and $\rho_\ell(s) = \xi (p^\star)^{2\ell +1} / 8\pi \sqrt{s}$ is the two-body phase space factor with barrier factors. Here, $\Theta$ is the Heaviside function.

Through the Schwartz reflection principle, one identifies the imaginary part of the amplitude with its discontinuities. The phase-space factor in Eq.~\eqref{eq:IV-PW-unitarity} gives a branch cut called the ``right-hand'' or ``unitarity'' cut. In general, amplitudes have both right-hand cuts (from direct channel unitarity, as shown) and left-hand cuts, which arise from cross-channel processes. Likewise, partial-wave amplitudes can have left-hand cuts induced from partial-wave projecting these cross-channel processes. To capture the generic analytic structure of the amplitude, we introduce a decomposition called $N/D$~\cite{Omnes:1958hv, Chew:1960iv, Bjorken:1960zz},
    \begin{align}
    \label{eq:ND_decomp}
    \tilde{\Mc}_\ell(s) \equiv \Nc_\ell(s) \, \Dc_{\ell}^{-1}(s) \, .
    \end{align}
Here, by definition, $\Dc_\ell(s)$ has only right-hand cuts. Zeros of $\Dc_\ell(s)$ on the physical sheet can only be on the real axis and correspond to bound states. Zeros in the complex energy plane on the unphysical Riemann sheet describe resonances or virtual states. The left-hand cuts lie below the physical region and are contained in the numerator function $\Nc_\ell(s)$. Therefore, the partial-wave unitarity condition separates into one condition for $\Dc_\ell(s)$,
    \begin{align}
    \im \Dc_\ell(s) = -\rho_\ell(s) \, \Nc_\ell(s) \, \Theta(s - s_{\mathrm{thr}}) \, , 
    \end{align}
and another expression for the imaginary part of $\Nc_\ell(s)$,
    \begin{align}
    \im \Nc_\ell(s) = \im \tilde{\Mc}_\ell(s) \, \Dc_\ell(s) \, \Theta(s_{\rm lhc} - s) \, ,
    \end{align}
where $s_{\rm lhc}$ is the location of the nearest left-hand cut. These conditions have known solutions in terms of coupled linear integral equations,
    \begin{align}
    \label{eq:D_int_eq}
    \Dc_\ell(s) & = 1 - \int_{s_{\mathrm{thr}}}^{\infty}\!\frac{\diff s'}{\pi} \, \frac{\rho_\ell(s')}{s' - s - i\epsilon} \, \Nc_\ell(s') \, , \\[5pt]
    \label{eq:N_int_eq}
    \Nc_\ell(s) & = \int_{-\infty}^{s_{\rm lhc}}\!\frac{\diff s'}{\pi} \, \frac{\im \tilde{\Mc}_\ell(s')}{s' - s - i\epsilon} \, \Dc_\ell(s') \, ,
    \end{align}
which are obtained from the Cauchy theorem under the assumption that $|\tilde{\Mc}_\ell(s)|$ and $|\Nc_\ell(s)| \to 0$ fast enough as $|s| \to \infty$. In particular, we assume $\Nc_\ell$ decays faster than $(p^\star)^{-2\ell}$ with $|s| \to \infty$, such that the integral in~\eqref{eq:D_int_eq} is convergent. In practice, various parametrizations of the amplitude do not satisfy this criterion, and subtracted dispersion relations must be used (or the behavior of the parametrization tuned at infinity) to eliminate a potential integral divergence and the ambiguity in the overall normalization of $\Nc$ and $\Dc$ functions (which can be simultaneously multiplied by any holomorphic function without affecting the amplitude.) We discuss it further in the supplemental material and here keep the simplest $N/D$ form for clarity of the presentation. Note that we assume that $\Dc$ is normalized according to $\Dc_{\ell}(\infty) = 1$.


\emph{Finite-volume scattering:} We now consider the constraint unitarity imposes on the system in a finite cubic volume of side length $L$, subjected to periodic boundary conditions so that momenta are quantized as $\p = 2\pi \n/L$ for $\n\in\mathbb{Z}^3$. The condition for the imaginary part of the FV amplitude $\Mc_L$ takes the form of Eq.~\eqref{eq:IV-unitarity} with the integral replaced by a sum performed over all discrete components of the intermediate momentum $\k$,
    \begin{align}
    \int \! \frac{\diff \k}{(2\pi)^3} \to \frac{1}{L^3} \sum_{\k} \, ,
    \end{align}
and the $\Mc \to \Mc_L$ switch. Importantly, it is assumed that the FV ground states have masses that for large enough volumes are exponentially close to the IV mass, $m_{i,L} = m_{i} + \Oc(e^{-m_\pi L})$. The finite-volume amplitude is decomposed into partial waves as,
    \begin{align}
    \label{eq:FV-PW}
    & \Mc_L(\p',\p; \P) = \\ \nonumber
    & 4\pi\, \sum_{\ell' m_\ell'} \sum_{\ell m_\ell} Y_{\ell' m'_\ell}(\hat{\p}'^\star) \, 
    \Mc_{L;\ell'm'_\ell,\ell m_\ell}(p'^\star, p^\star; \P) \, 
    Y_{\ell m_\ell}^*(\hat{\p}^\star) \, .
    \end{align}
The above expression has to be treated with caution. Since rotational symmetry is broken in the box, spherical harmonics no longer form a set of orthonormal, linearly independent states of well-defined angular momentum. Thus, it is difficult to unambiguously define partial-wave projection and its inverse for discrete amplitudes~\cite{Li:2019qvh, Miyamoto:2019jjc}. $\Mc_L(\p',\p;\P)$ appearing in \eqref{eq:FV-PW} has to be understood as an extension of the half-off-shell amplitude from discrete to continuous momenta. We provide a more thorough discussion of the finite-volume amplitudes, their properties, and the definition of the partial waves in the supplemental material.

Since rotational symmetry is broken, one cannot reduce the expansion further. The FV amplitude is a function of $\P$ and not just $s$ (or equivalently $p'^\star = p^\star$), as the scattering depends on the orientation of $\P$ with respect to the box. The partial-wave unitarity relation becomes,
\begin{widetext}
    \begin{align}
    \label{eq:FV-PW-unitarity}
    2\im \Mc_{L}(p^\star, \P) = & \frac{\xi}{L^3}\sum_{\k} \,  \frac{\pi \delta(E - \omega_1(\k) - \omega_2(\P-\k)) }{2\omega_1(\k) \, 2\omega_2(\P-\k)} \, (4\pi) \,
    \left[
    Y^{\, \dag}(\hat{\k}^\star) \, 
    \Mc_{L}(p^\star, \P)
    \right]^\dag
    \otimes
    \left[
    Y^{\, T}(\hat{\k}^\star) \,
    \Mc_{L}(p^\star, \P) 
    \right] \, ,
    \end{align}
\end{widetext}
where we introduced a notation in which amplitudes are considered matrices in the $(\ell'm_\ell'; \ell m_\ell)$ space; for instance $[\Mc_L(p^\star, \P)]_{\ell' m_\ell'; \ell m_\ell} = \Mc_{L;\ell' m_\ell'; \ell m_\ell}(p^\star, p^\star; \P)$. Spherical harmonics become vectors in this notation, $[Y(\hat{\k}^\star)]_{\ell m_\ell} = Y_{\ell m_\ell}(\hat{\k}^\star)$. We note that under the sum, $k^\star = p^\star$ due to the Dirac delta function and both momentum magnitudes are fixed by the invariant mass $s$. Similarly to the infinite volume, we remove the threshold barrier factors from the amplitude by defining,
    \begin{align}
    \Mc_{L}({p}^\star, \P) = Q(p^\star) \, \tilde{\Mc}_{L}({p}^\star, \P) \, Q(p^\star) \, ,
    \end{align}
where the matrix $[Q(p^\star)]_{\ell' m_\ell' ; \ell m_\ell} = \delta_{\ell' \ell} \, \delta_{m_\ell' m_\ell} \, (p^\star)^{\ell}$. The analytic structure of $\tilde{\Mc}_{L}(p^\star, \P)$ is determined by a collection of simple right-hand poles in the $E^\star$ variable originating from the direct channel. There are also finite-volume bound-state poles and the cross-channel cuts or/and poles that are considered as left-hand singularities. We can then write the $N/D$ decomposition for the FV amplitude,
    \begin{align}
    \label{eq:FV-ND}
    \tilde{\Mc}_{L}({p}^\star, \P) = \Nc_{L}(p^\star, \P) \, \big(\Dc_L(p^\star, \P) \big)^{-1}  \, ,
    \end{align}
in which the IV $\Nc$ and $\Dc$ functions become matrices in angular momentum space. Note that the partial-wave mixing in the finite volume effectively turns the FV $N/D$ representation into a ``coupled channel'' problem. We stress that by construction, the FV energy levels (right-hand and bound-state poles of the amplitude) are given solely by zeros of $\Dc_L$ and not by the potential left-hand poles in $\Nc_L$. These two functions are related to each other, and any right-hand pole introduced in a given parametrization of $\Nc_L$ can be replaced by a zero of $\Dc_L$ without affecting the scattering amplitude $\tilde{\Mc}_L$. This is obtained, for instance, by multiplying $\Nc_L$ and $\Dc_L$ by a holomorphic function with zeros coinciding with the relevant poles of $\Nc_L$.

By demanding that $\Nc_L$ contains the left-hand singularities only, $\im \Nc_L(p^\star) = 0$ in the physical direct channel, and therefore performing matrix manipulations analogous to those in Ref.~\cite{PhysRev.121.1250}, we find,
\begin{widetext}
    \begin{align}
    \label{eq:FV-Im-D}
    \im \Dc_{L}(p^\star, \P) & = 
    - \frac{\xi}{L^3} 
    \sum_{\k} \, 
    \frac{2s}{E} \, 
    \frac{\pi \delta \big(s - E^\star(\k)^2 \big)}
    {2\omega_1(\k) 2 \omega_2(\P - \k)} \, 
    \Yc^*(\k^\star) 
    \otimes \left( 
    \Yc(\k^\star)^T \,
    \Nc_{L}(k^\star, \P) 
    \right) \, ,
    \end{align}
\end{widetext}
where we introduced $E^\star(\k) = \omega_1(\k^\star) + \omega_2(\k^\star)$ and $\Yc(\k^\star) = \sqrt{4\pi} \, Q(k^\star) \, Y(\hat{\k}^\star)$, and rewrote the Dirac delta as a function of the Mandelstam $s$ variable. It should be understood that $p^\star$ is considered a function of $s$, as given in Eq.~\eqref{eq:momentum}. The solution to the unitarity equation for $\Dc_L$ is given by a sum of poles,
\begin{widetext}
    \begin{align}
    \Dc_{L}(p^\star,\P) & = \one + \int_{s_{\rm thr}}^{\infty} \! \frac{\diff s'}{\pi} \, \frac{\im \Dc_{L}(q^\star,\P)}{s' - s - i\epsilon } = \one - \frac{\xi}{L^3} \sum_{\k} 
    \Lc(\k,\P) \, 
    \frac{ 
    \Yc^*(\k^\star) 
    \otimes 
    \Yc(\k^\star)^T }
    {E^\star(\k)^2 - s - i\epsilon} \, 
    \Nc_{L}(k^\star, \P) \, ,
    \label{eq:d_L_dispersive-1}
    \end{align}
\end{widetext}
where $q^\star$ is a momentum corresponding to the $s'$ variable and we define the kinematic factor,
    \begin{align}
    \Lc(\k,\P) = \frac{E^\star(\k)^2}{2\omega_1(\k) \, 2\omega_2(\P-\k) \, \sqrt{ E^\star(\k)^2 + \P^2 }} \, .
    \end{align}

Three comments are in order. Firstly, let us note that the energy denominator becomes zero (leads to a $\Dc_L$ pole) at free energies $s = E^\star(\k)^2$. Thus, just like the finite-volume amplitude, $\Dc_L$ considered a function of $s = (E^\star)^2$ has a non-zero imaginary part above the threshold. Since one is interested in the interacting energies---corresponding to zeros and not poles of $\Dc_L$---the $i\epsilon$ can be safely removed from the denominator. This holds only if the interacting energy levels do not coincide with the free ones. Furthermore, it is assumed that $\Nc_L$ decays fast enough for large momenta, such that the sum in Eq.~\eqref{eq:d_L_dispersive-1} is convergent. If this cannot be ensured, one needs to regulate the sum by including subtractions, as discussed in the supplemental material.

Finally, we note that $\Nc_L$ has no singularities in the physical momentum region, over which the sum is performed. Therefore, the FV numerator function can be replaced in the above expression by its IV version, $\Nc$, if one neglects corrections that are exponentially suppressed with the size of the box. In this context, $\Nc$ is considered a diagonal matrix with elements given by $[\Nc]_{\ell'm_\ell';\ell m_\ell} = \delta_{\ell' \ell} \, \delta_{m_\ell m_\ell'} \, \Nc_{\ell}(k^\star)$, where we wrote $\Nc_\ell$ as a function of $k^\star$, $ \Nc_\ell(s') \equiv \Nc_\ell(k^\star)$. A detailed argument for this replacement is provided in the supplemental material. 

\begin{widetext}
Given this replacement, one can write,
    \begin{align}
    \Dc_{L}(p^\star,\P) = \one + \frac{\xi}{L^3} \sum_{\k} 
    \Lc(\k,\P) \, 
    \frac{\Yc^*(\k^\star)
    \otimes 
     \Yc(\k^\star)^T }
    {s - E^\star(\k)^2} \, 
    \Nc(k^\star) \, ,
    \label{eq:d_L_dispersive}
    \end{align}
\end{widetext}
and the eigenenergies of the FV theory are given by,
    \begin{align}
    \label{eq:ND_QC}
    \det_{\ell m_\ell} \Dc_L = 0 \, ,
    \end{align}
where the determinant is taken in the angular momentum space $(\ell,m_\ell)$. Therefore---just like the L\"uscher quantization condition---Eq.~\eqref{eq:ND_QC} relates the spectrum of the theory in a spatial box to the IV scattering information; in this case the numerator function $\Nc_{\ell}(s)$. Unlike in the derivation of the L\"uscher QC, at no point were we required to assume $s > s_{\rm lhc}$, and the provided formula applies at energies below the nearest left-hand cut branch point. The key difference is that our formalism involves only the $\Dc$ function, while the L\"uscher's approach involves both $\Nc$ and $\Dc$ through the K matrix. This does not affect the elastic energy region, where both approaches are equivalent, as we demonstrate in the supplemental material. However, for energies below the threshold, the model-independent separation of the singularities between $\Dc$ and $\Nc$ allows one to determine the spectrum of the FV theory without evaluating singular parts of the amplitude from the cross-channel processes, which leads to unaccounted power-law volume dependence in na\"ive extrapolation of L\"uscher QC.

We stress that the $N/D$ form, just like the commonly used $K$ matrix parametrization of partial waves is a generic description of scattering amplitude satisfying the S-matrix unitarity and, additionally, incorporating analyticity through the dispersion relations. Given suitable parametrizations of $\Nc_{\ell}(s)$, that could in principle incorporate one- or multi-particle exchanges, one can constrain the numerator function using lattice data, and then reproduce the full solution by using the IV $N/D$ dispersive integrals given by Eqs.~\eqref{eq:D_int_eq} and~\eqref{eq:N_int_eq}. For the practitioner's convenience, we discuss examples of such parametrizations in the supplement.


\emph{Conclusions: } We derived a novel form of the two-body quantization condition formulated through the S-matrix principles of unitarity and analyticity. It incorporates physics from left-hand cuts stemming from particle exchange processes by employing the $N/D$ formalism. It separates the left-hand from right-hand cuts in the FV analog of the scattering amplitude and allows one to express the quantization condition entirely in terms of the FV $\Dc_L$ matrix. It circumvents the restrictions from the original L\"uscher condition, valid only for energies greater than the nearest left-hand branch point. The $\Dc_L$ denominator is related to the numerator $\Nc$; thus, one may use the lattice QCD spectrum to constrain its parametrizations. These may be built from the known locations of left-hand branch points---independent of the local interactions---and residual flexible models describing short-range physics. After the parameters of $\Nc$ are fitted from available data, one can reconstruct the full partial-wave amplitude from the $N/D$ integral relations.

We hope the idea presented in this work will become a starting point for further fruitful generalizations and numerical applications. Future extensions of the formalism must consider coupled-channel systems and particles with non-zero spin, for which known issues with the symmetry of $N/D$ equations lead to subtleties in building suitable parametrizations of the amplitude~\cite{PhysRev.121.1250, Martin:1963zz, Badalian:1981xj, Oller:2019opk}. Moreover, the formal investigation of angular momentum basis truncation in Eq.~\eqref{eq:ND_QC} has to be performed for specific models of $\Nc$. Finally, a formal comparison of our formalism with other available approaches~\cite{Klos:2016fdb, Meng:2023bmz, Raposo:2023oru, Bubna:2024izx, Hansen:2024ffk} would be desirable. From the practical side, after appropriate generalizations the potential applications of the presented formalism may include reactions like $NN$ or the $DD^*$ scattering, relevant for the study of the $T_{cc}^+$ tetraquark~\cite{Padmanath:2022cvl, Lyu:2023xro, Chen:2022vpo, Ortiz-Pacheco:2023ble, Du:2023hlu, Meng:2023bmz, Collins:2024sfi, Hansen:2024ffk, Whyte:2024ihh, Abolnikov:2024key}.


\emph{Acknowledgements:} We thank Marshall Baker, Jeremy Green, and Stephen Sharpe for useful discussions. We also thank Max Hansen, Robert Perry, Akaki Rusetsky, and Stephen Sharpe for their valuable comments on the manuscript.

SMD acknowledges the financial support through the U.S. Department of Energy Contract No. DE-SC0011637. This work was supported by the U.S. Department of Energy contract DE-AC05-06OR23177, under which Jefferson Science Associates, LLC operates Jefferson Lab, by the U.S. Department of Energy Grant No. DE-FG02-87ER40365, and contributes to the aims of the U.S. Department of Energy ExoHad Topical Collaboration, contract DE-SC0023598.

\bibliographystyle{apsrev4-1}
\bibliography{main}

\clearpage

\onecolumngrid

\section*{supplemental material}

In this supplement, we briefly describe the more technical aspects of the presented work. In part~\ref{app:formal}, we discuss the definition of the amplitude through the finite-volume scattering operator. In part~\ref{app:subtractions}, we discuss the importance of subtractions in the $N/D$ equations for the new quantization condition. In part~\ref{app:n-is-nl}, we show that the numerator functions in infinite and finite volume match up to corrections which scale as $O(e^{-\Delta L})$, where $\Delta$ is the lowest energy scale of the problem, e.g., the pion mass. In part~\ref{app:Luscher}, we prove that our formalism reduces to L\"uscher’s condition in the elastic energy region. Finally, in part~\ref{app:n-params}, we discuss simple parametrizations of the IV numerator function, $\Nc$, applicable in practical computations of physical systems.

\subsection{Formal aspects of the FV amplitudes}
\label{app:formal}

The FV transition operator is defined in complete analogy to its IV version. Namely, we define a theory in a box, $H_L = H_{0,L} + V_{L}$, where $H_{0,L}$ is the free and $V_{L}$ is the interaction Hamiltonian. The transition operator is,
    \begin{align}
    \label{eq:t-operator}
    T_L(z) = V_L + V_L \, G_L(z) \, V_L \, ,
    \end{align}
where $G_L(z) = (z-H_L)^{-1}$ is the Hamiltonian's resolvent (Green's operator.) By construction, it inherits the singularity structure of $G_L(z)$, e.g., the poles of $T_L(z)$ in the $z$ variable coincide with the spectrum of the theory. The main difference between the FV and IV theories comes from the spaces of scattering states they are built upon. In the FV case, periodic boundary conditions mean that the three-dimensional coordinate space is topologically equivalent to a torus. Its size is assumed to be much larger than the range of interactions, such that the asymptotic behavior of the hadrons is affected only by the boundary conditions. The finite-volume space of states consists of quantized momentum kets, $|\p \rangle$, where $\p = 2\pi \n/L$, $\n \in \mathbb{Z}^3$. They are normalized to Kronecker deltas, while the IV states correspond to a continuous momentum spectrum and are normalized to Dirac deltas. The set of FV scattering states can be given a well-defined meaning of a Hilbert space, while such definition in the infinite volume is accomplished in the sense of the rigged Hilbert space~\cite{gel2014generalized, Maurin, de_la_Madrid_2005}.

The half-off-shell FV scattering amplitude, $\Mc_L$, is defined as the matrix element of $T_L(z)$ with $z = E + i 0^+$, taken between (discrete) momentum eigenstates of hadrons moving inside the torus. Starting from the definition of the FV transition operator and following the same derivation method as in the infinite volume, one finds that the imaginary part of ${\cal M}_L$ satisfies the off-shell unitarity relation,
    \begin{align}
    \label{eq:FV-unitarity}
    2\im \Mc_{L}(\p',\p; \P) = \frac{\xi}{L^3}\sum_{\k} \,  \frac{\pi \delta(E - \omega_1(\k) - \omega_2(\P-\k)) }{2\omega_1(\k) \, 2\omega_2(\P-\k)} \, \Mc_{L}^*(\k,\p'; \P)  \, \Mc_{L}(\k,\p ; \P) \, .
    \end{align}
Certain care is required when interpreting this relation. In particular, unlike in the infinite volume, this equation alone cannot be used to infer the singularity structure of the amplitude without additional input, e.g., from Eq.~\eqref{eq:t-operator}. The FV amplitude has poles at interacting energies that occur above the threshold, which implies 
that its imaginary part is non-zero outside of the set of free energies, $\omega_1(\k) + \omega_2(\P-\k)$, despite being na\"ively canceled by the explicit vanishing Dirac delta in Eq.~\eqref{eq:FV-unitarity}. This can be inferred from the operator relation $T_L(z) - T_L(z^*) = 2\pi i \, V_L \, \delta(z - H_L) \, V_L = 2 \pi i \, T_L(z) \, \delta(z - H_{0,L}) \, T_L(z^*)$ obtained from the first resolvent identity. Moreover, as can be seen directly from the definition of the transition operator, the FV amplitude vanishes at free energies (assuming they are separated from the interacting energies), removing the explicit contributions from the free-energy Dirac deltas in the unitarity equation. Nonetheless, as described in the main text, one may manipulate the FV unitarity relation into a simpler equation that describes singularities of its inverse, i.e., the equation for the imaginary part of the $\Dc_L$ function, Eq.~\eqref{eq:FV-Im-D}, which exhibits intuitive analytic properties (poles at free energies and zeros at interacting energies.) This is analogous to the situation in the infinite volume, where the unitarity equation for the inverse of the amplitude gives its right-hand discontinuity, $\im \tilde \Mc_\ell^{-1}(s) = - \rho_\ell(s)$.

Because the set of relative momenta and their orientations is discrete, spherical harmonics no longer define orthonormal, linearly independent states of well-defined angular momentum, $| \ell, m_\ell \rangle$. To define the partial-wave projection of $\Mc_L(\p',\p;\P)$, we extend it to continuous momenta. It is done by observing that although the FV and IV interaction Hamiltonian operators are different mathematical entities due to the distinct domains they are defined for, their momentum-space matrix elements are identical\footnote{Up to trivial momentum-conserving Kronecker and Dirac deltas.} and one can write,
    \begin{align}
    \label{eq:extended}
    \Mc_L(\p',\p; \P) = V(\p',\p) + \sum_{\{\q'\}_i} \sum_{ \{\q\}_j} V(\p', \{\q'\}_i) \, G_L(\{\q'\}_i, \{\q\}_j; z) \, V(\{\q\}_j, \p) \, .
    \end{align}
Here, the intermediate sums are performed over all multi-particle sectors of the Fock space (labeled by indices $i$ and $j$), and all discrete momenta $\{\q'\}_i = \{\q'_1, \q'_2, \dots, \q'_i\}$ and $\{\q\}_j = \{\q_1, \q_2, \dots, \q_j\}$ of particles belonging to these sectors that are consistent with the total three-momentum conservation. For clarity, we absorbed all normalization and $1/L^3$ factors into a definition of the sum. $A(\{ \p\}_i,\{\k\}_j)$ is a matrix element of a generic operator $A$ between multi-particle states $| \{\p\}_i \rangle$ and $|\{\k\}_j\rangle$ from sectors $i$ and $j$, respectively. We represent two-body subspaces just with the value of the relative momentum, $\{\p\}_2 = \{\p_1,\p_2\} \equiv \p$ (that is, the amplitude is considered a function of the arguments $\p'$, $\p$, and $\P$ rather than $\p_1'$, $\p_2'$, $\p_1$, and $\p_2$.) Equation~\eqref{eq:extended} defines $\Mc_L(\p',\p; \P)$ for arbitrary momenta, including those of continuous and complex magnitudes. Given this expression, for every theory in a box, a clear meaning can be assigned to the partial waves, $\Mc_{\ell' m', \ell m}(p'^\star, p^\star; \P)$, appearing in Eq.~\eqref{eq:FV-ND},
    \begin{align} 
    \Mc_L(\p',\p; \P) = 
    \sum_{\substack{\ell' m' \\ \ell m}} Y_{\ell' m'}(\Omega_{p'})  \left( V_{\ell' m; \ell m}(p'^\star,p^\star) 
    + 
    \!\!\! \sum_{\{\q'\}_i, \{\q\}_j} \!\!\!
    V_{\ell' m'}(p'^\star, \{\q'\}_i) \, 
    G_L(\{\q'\}_i, \{\q\}_j; z) \, 
    V_{\ell m}(\{\q\}_j, p^\star) \right) 
    Y^*_{\ell m}(\Omega_p) \, ,
    \end{align}
where $V$ with discrete indices are appropriate partial-wave projections of the matrix elements of the IV interaction Hamiltonian. We do not apply this equation in practice, but it serves as a formal definition of the partial-wave projected amplitude in a box. Using the decomposition of the amplitude in Eq.~\eqref{eq:FV-unitarity} leads to Eq.~\eqref{eq:FV-PW-unitarity}.

Given this definition of the partial-wave projection, in a generic quantum field theory with local interactions, partial-wave projection of $V$ matrix elements produces smooth functions of momenta. Left-hand singularities are given by the matrix elements of Green's function evaluated at energies corresponding to discrete momentum transfers between two short-range couplings and take the form of left-hand poles. It is interesting to note that the left-hand cuts can also originate in the finite volume. This happens in systems governed by long-range, non-local interactions such as those described by Yukawa-type potentials in non-relativistic quantum mechanics.

Thus, in addition to the unitarity poles (and in agreement with na\"ive expectation from crossing symmetry), the FV partial wave amplitude inherits singularities from the crossed channels. Since these are also constrained by unitarity, one indeed expects poles in the FV instead of cuts when the amplitude is considered a function of the cross-channel Mandelstam variables in relativistic theory. The dispersion relations for the $\Nc_L$ and $\Dc_L$ functions are proposed based on this singularity structure.

\subsection{Subtracted $N/D$ equations}
\label{app:subtractions}

To ensure (or improve) convergence of the $N/D$ integrals, it is common to introduce subtractions. One writes,
    \begin{align}
    \label{eq:D_int_eq1}
    \Dc_\ell(s) & = d_\ell^{(j)}(s) - (s-s_0)^j \int_{s_{\mathrm{thr}}}^{\infty} \! \frac{\diff s'}{\pi} \, \frac{\rho_\ell(s')}{(s' - s - i\epsilon) (s' - s_0)^j } \, \Nc_\ell(s') \, , \\
    \label{eq:N_int_eq1}
    \Nc_\ell(s) & = n_{\ell}^{(i)}(s) + (s-s_0)^i \int_{-\infty}^{s_{\rm lhc}}\!\frac{\diff s'}{\pi} \, \frac{\im \Mc_\ell(s')}{(s' - s - i\epsilon) (s' - s_0)^i } \, \Dc_\ell(s') \, ,
    \end{align}
where $i,j \in \mathbb{N}$ and $n_\ell^{(i)}$, $d_\ell^{(j)}$ are some polynomials of the $i$th and $j$th order, respectively. The choice of the subtraction point $s_0$ redefines $\Nc$ and $\Dc$ by a constant multiplicative factor. In many cases of practical interest, the simplest, single subtraction in the dispersive integral for $\Dc_\ell$ is enough to guarantee convergence of the $N/D$ representation. Focusing on such an example, we write,
    \begin{align}
    \Dc_\ell(s) & = 1 - \int_{s_{\mathrm{thr}}}^{\infty} \! \frac{\diff s'}{\pi} \, \frac{s}{s'} \, \frac{\rho_\ell(s')}{(s' - s - i\epsilon)} \, \Nc_\ell(s') \, , \\
    \label{eq:N_int_eq2}
    \Nc_\ell(s) & = \int_{-\infty}^{s_{\rm lhc}}\!\frac{\diff s'}{\pi} \, \frac{\im \Mc_\ell(s')}{s' - s - i\epsilon } \, \Dc_\ell(s') \, ,
    \end{align}
where we picked $s_0 = 0$. Following the logic of the main text, it can be shown that the corresponding quantization condition is given by Eq.~\eqref{eq:ND_QC} with,
    \begin{align}
    \Dc_{L}(p^\star,\P) = \one - \frac{\xi}{L^3} \sum_{\k} 
    \frac{s \, \Lc(\k,\P) }{E^\star(\k)^2} \, 
    \frac{\Yc^*(\hat{\k}^\star)
    \otimes 
     \Yc(\hat{\k}^\star)^T }
    {s - E^\star(\k)^2} \,
    \Nc(k^\star) \, .
    \label{eq:d_L_dispersive1}
    \end{align}
Note that the additional factor does not affect the proof that $\Nc = \Nc_L$ for physical kinematics presented in part~\ref{app:n-is-nl}, nor the equivalence of our formalism with the L\"uscher quantization condition derived in part~\ref{app:Luscher}. This is because the factor $s/E^\star(\k)^2$ can be replaced by $1$ under every sum-integral difference appearing in these derivations.

\subsection{Proof that $\Nc = \Nc_L + \Oc(\exp(- \Delta L))$}
\label{app:n-is-nl}

\begin{figure}[h!]
    \centering
    \includegraphics[width=0.90\textwidth]{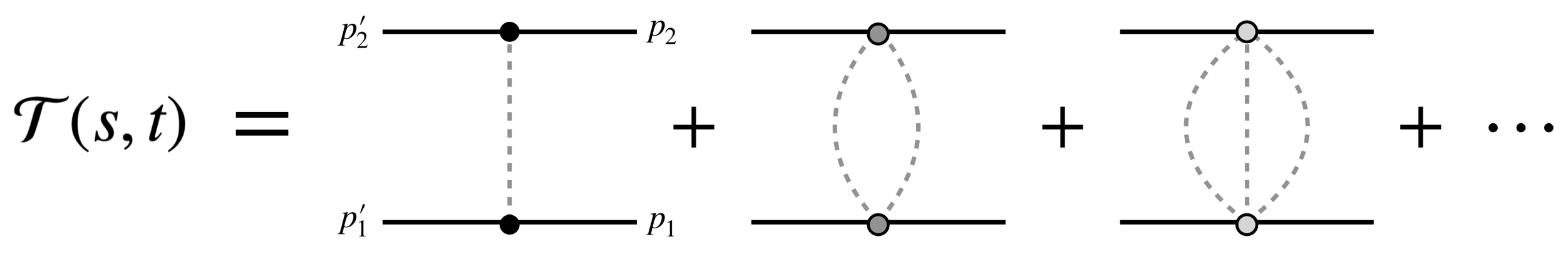}
    \caption{Example Feynman diagrams contributing to the left-hand singularity structure of the partial-wave amplitude in some generic effective field theory. Since particles in loops can not go on-shell for the physical energies, these diagrams do not lead to the right-hand singularities.}
    \label{fig:diags}
\end{figure}

Let us first consider the IV partial-wave amplitude $\Mc = \Nc \Dc^{-1}$, where each object is a diagonal matrix in the $(\ell' m_\ell'; \ell m_\ell)$ space. For simplicity, we assume that the left-hand singularities of the amplitude $\Mc$ are given by one- and multi-particle exchange diagrams as shown in Fig.~\ref{fig:diags}. It is not necessary for the derivation but allows us to focus on an example of physical interest. Contributions for such terms are denoted collectively by $\Tc(s,t)$. The $\ell$-th partial-wave of $\Tc(s,t)$ in the $s$ channel is called $\Tc_{\ell}(s)$ and is a matrix element of the (diagonal) matrix $\Tc(s)$. Since $\im \Mc(s) = \im \Tc(s)$ for $s < s_{\rm lhc}$ and  $\im \Tc(s) = 0$ for $s>s_{\rm thr}$, one may write,
    \begin{align}
    \Nc(s) & = \int_{-\infty}^{s_{\rm lhc}} \frac{\diff s'}{\pi} \frac{\im \Tc(s')}{s' - s - i \epsilon} 
    - \int_{-\infty}^{s_{\rm lhc}} \frac{\diff s'}{\pi} \frac{\im \Tc(s') }{s' - s - i \epsilon} 
    \int_{s_{\rm thr}}^{\infty} \frac{\diff s''}{\pi} \frac{\rho_\ell(s'') \, \Nc(s'') }{s'' - s' - i \epsilon} \\
    & = \Tc(s) 
    + \int_{s_{\rm thr}}^{\infty} \frac{\diff s''}{\pi} \frac{\rho_\ell(s'') \, \im \big[ \Tc(s'') -  \Tc(s) \big] }{s'' - s - i \epsilon} \, \Nc(s'') \, , 
    \label{eq:proof1}
    \end{align}
where we have used the integral equation for $\Dc_\ell$ in the formula for $\Nc_\ell$, and integrated over $s'$ in the second line. As a result, we obtain a single integral equation for $\Nc_\ell(s)$, which can be solved once $\Tc$ is specified. In the following, we assume that all integrals converge. Rigorously, it is incorrect for the type of $\Tc$ contributions considered here (which lead to logarithmic divergences in the $N/D$ equations.) Divergences are taken care of by introducing subtractions, as discussed in~\ref{app:subtractions}, and we proceed without them to maintain the simplicity of the expressions in this argument.

In the FV theory, the $\Nc_L$ is obtained from dispersing its imaginary part $\im \Nc_L = \im \Mc_L \, \Dc_L$. Thus, taking $\Dc_L$ given by Eq.~\eqref{eq:d_L_dispersive-1} and neglecting exponentially suppressed corrections, we obtain,
    \begin{align}
    \Nc_L(p^\star,\P) = \Tc(s) 
    + \frac{1}{L^3} \sum_{\k} \, \Lc(\k, \P) \, \frac{ \left[\Yc^*(\k^\star) \otimes \Yc(\k^\star)^T \right] \im 
    \big[ \Tc(E^\star(\k)^2) -  \Tc(s) \big]}{E^\star(\k)^2 - s - i \epsilon} \, \Nc_L(k^\star, \P) \, .
    \label{eq:proof2}
    \end{align}
As usual, $p^\star$ is the momentum corresponding to the total invariant mass $s$. To obtain the above expression we performed similar operations as in Eq.~\eqref{eq:proof1}, assumed $p^\star > 0$ (physical kinematics,) and used the fact that $\Tc_L(p^\star) = \Tc(p^\star) + \Oc(e^{-\Delta L})$, where $\Delta$ represents the lowest energy scale in the problem (e.g., pion mass.) 

Note that neither $\Nc_L$ nor $\Tc$ have singularities for physical momenta over which one performs the sum in Eq.~\eqref{eq:proof2}. Moreover, the difference of two $\Tc$'s in the numerator regulates the pole singularity from the vanishing energy denominator. Therefore, by the Poisson summation formula, the sum may be replaced by the corresponding IV integral over momenta---up to the corrections exponentially suppressed with the box size. A similar argument is used to show that~\eqref{eq:appc1} is equal to~\eqref{eq:appc2} in~\ref{app:Luscher}. By variable transformation, one rewrites it as an integral over $s'$, identical to the one appearing in Eq.~\eqref{eq:proof1}. It shows that $\Nc = \Nc_L + \Oc(e^{-\Delta L})$ in the physical energy region, which was necessary for the derivation of the $N/D$ quantization condition in the main text, Eq.~\eqref{eq:d_L_dispersive}.

\subsection{Reduction to the L\"uscher formalism} 
\label{app:Luscher}

The L\"uscher quantization condition can be expressed as,
    \begin{align}
    \det_{\ell m_\ell} \Big[\Kc^{-1}(s) + F(P,L)\Big] = 0 \, ,
    \label{eq:luscher}
    \end{align}
which holds for the FV spectrum for a given momentum $\P$ and box size $L$. Here, $\Kc$ is the IV two-body $K$ matrix, $[\Kc]_{\ell' m_\ell'; \ell m_\ell} = \delta_{\ell' \ell} \, \delta_{m_\ell' m_\ell} \, \Kc_{\ell}(s)$, related to the scattering amplitude via,
    \begin{align}
    \Mc_{\ell}^{-1}(s) = \Kc_\ell^{-1}(s) - i \rho(s) \, ,
    \end{align}
where $\rho$ is the phase space factor without barrier factors, $\rho(s) = \xi p^\star / 8\pi \sqrt{s}$.
Finally, $F(P,L)$ is a matrix of elements~\cite{Briceno:2015tza},
    \begin{align}
    F_{\ell' m_\ell' ; \ell m_\ell}(P,L) = \xi \left[\frac{1}{L^3}\sum_{\k} - {\rm p.v.} \int\! \frac{\diff \k}{(2\pi)^3}\right] \,
    \left(\frac{k^\star}{p^\star}\right)^{\ell'} 
    \frac{(4\pi) \, 
    Y^*_{\ell'm_{\ell'}}({\hat{\k}}^\star) \,
    Y_{\ell m_{\ell}}(\hat{\k}^\star)}
    {2\omega_1(\k) \, 2 \omega_2(\P-\k) \,  \big( E - \omega_1(\k) - \omega_2(\P-\k) \big)} 
    \left(\frac{k^\star}{p^\star}\right)^{\ell} \, ,
    \end{align}
where ``p.v.'' indicates the Cauchy principal value prescription. 

To compare our and L\"uscher's quantization condition, several tedious manipulations are necessary. First, we note that the $N/D$ representation introduced in the main text relates the spectrum to the amplitude $\tilde{M}$, stripped off the threshold barrier factors,
    \begin{align}
    \tilde{\Mc}_{\ell}^{-1}(s) = \tilde{\Kc}_\ell^{-1}(s) - i \rho_\ell(s) \, ,
    \end{align}
where $\tilde{\Kc}_\ell = (p^\star)^{2\ell} \,  \Kc_\ell$. Removing these factors from the quantization condition gives, 
    \begin{align}
    \det_{\ell m_\ell} \Big[\tilde{\Kc}^{-1}(s) + \tilde{F}(P,L)\Big] = 0 \, ,
    \label{eq:luscher1}
    \end{align}
where,
    \begin{align}
    \tilde{F}_{\ell' m_\ell' ; \ell m_\ell}(P,L) = \xi \left[\frac{1}{L^3}\sum_{\k} - {\rm p.v.} \int\! \frac{\diff \k}{(2\pi)^3}\right] \, 
    \frac{(4\pi) \, 
    (k^\star)^{\ell'} \, 
    Y^*_{\ell'm_{\ell'}}({\hat{\k}}^\star) \,
    Y_{\ell m_{\ell}}(\hat{\k}^\star) \, 
    ( k^\star)^{\ell} }
    {2\omega_1(\k) \, 2 \omega_2(\P-\k) \, 
    \big(E - \omega_1(\k) - \omega_2(\P-\k) \big)} \, .
    \end{align}
In the next step, we modify the energy denominator of the above function. We add and subtract $(E + \omega_1(\k) + \omega_2(\P-\k))^{-1}$ to write,
    \begin{align} 
    \frac{1}
    {E - \omega_1(\k) - \omega_2(\P-\k)} = \frac{2E}
    {E^2 - (\omega_1(\k) + \omega_2(\P-\k))^2  } - \frac{1}
    {E + \omega_1(\k) + \omega_2(\P-\k)} \, ,
    \end{align}
under the sum-integral difference. The contribution from the second term is exponentially suppressed with the box size, $L$, as it does not contain singularities in the elastic region. We drop it and use the c.m.~variables in the numerator, replacing, $E^2 - (\omega_1(\k) + \omega_2(\P-\k))^2 = \gamma^2 \big( s -  (\omega_1(\k^\star) + \omega_2(\k^\star))^2 \big)$, where $\gamma = E/E^\star$,
    \begin{align}
    \frac{2E}
    {E^2 - (\omega_1(\k) + \omega_2(\P-\k))^2  } = 2 \, \frac{(E^\star)^2}{E} \, \frac{1}
    {s - (\omega_1(\k^\star) + \omega_2(\k^\star))^2  } \, .
    \end{align}
Finally, we add and subtract $(\omega_1(\k^\star) + \omega_2(\k^\star))^2/\sqrt{(\omega_1(\k^\star) + \omega_2(\k^\star))^2 - \P^2}$ from the first fraction. Again, this results in an expression in which some contributions cancel the pole in the second fraction and, thus, are exponentially suppressed with volume under the sum-integral difference. Neglecting them, one can write,
    \begin{align}
    \tilde{F}(P,L) = \xi \left[\frac{1}{L^3}\sum_{\k} - {\rm p.v.} \! \int\! \frac{\diff \k}{(2\pi)^3}\right] \,
    \Lc(\k,\P) \,
    \frac{ 
    \Yc^*({\k}^\star) 
    \otimes
    \Yc(\k^\star)^T }
    {s - E^\star(\k)^2}  \, ,
    \label{eq:tilde_F}
    \end{align}
where we used the notation introduced in the main text. 

We now compare the IV $K$ matrix and $N/D$ parametrization for $s > s_{\rm thr}$. We start from,
    \begin{equation}
    \tilde{\Kc}^{-1}(s) - i \, Q(p^\star) \, \rho(s) \, Q(p^\star) = \Dc(s) \, \Nc^{-1}(s) \, .
    \end{equation}
where we consider all the IV objects as diagonal matrices in the $(\ell' m_\ell; \ell m_\ell)$ space. The $K$ matrix is given by the real part of the integral representation of $\Dc$, Eq.~\eqref{eq:D_int_eq},
    \begin{equation}
    \tilde{\Kc}^{-1}(s) = \left[ \one - {\rm p.v.} \! \int \frac{ \diff s'}{\pi} \, \frac{Q(k^\star) \rho(s') \, Q(k^\star) \, \Nc(s') }{s' - s} \right] \, \Nc^{-1}(s) \, ,
    \end{equation}
where $k^\star$ is a c.m.~momentum corresponding to the integration variable $s'$. We change the integration variables to momenta, adding a redundant angular part,
    \begin{align}
    \tilde{\Kc}^{-1}(p^\star) = \left[ \one - \xi \, {\rm p.v.} \! \int \frac{\diff \k^\star}{(2\pi)^3} \, \Lc(\k,\P) \, \frac{ \Yc^*(\k^\star) \otimes \Yc(\k^\star)^T \Nc(k^\star) }{E^\star(\k)^2 - s} \right] \, \Nc^{-1}(p^{\star}) \, .
    \end{align}
Here, we expressed the $\Kc$ and $\Nc$ as functions of momenta. We use the above representation and Eq.~\eqref{eq:tilde_F} in the L\"uscher quantization condition. Multiplying by $\det \Nc(p^\star)$ yields,
    \begin{align}
    \nonumber
    \det_{\ell m_\ell} \Bigg[ & \one + \xi \, {\rm p.v.} \! \int \frac{\diff \k}{(2\pi)^3} \, \Lc(\k,\P) \, \frac{\Yc^*(\k^\star) \otimes \Yc(\k^\star)^T }{s - E^\star(\k)^2} \, \Nc(k^\star) \\
     & + 
     \xi \left[\frac{1}{L^3}\sum_{\k} - {\rm p.v.} \! \int\! \frac{\diff \k}{(2\pi)^3}\right] \,
    \Lc(\k,\P) \,
    \frac{\Yc^*(\k^\star) \otimes \Yc(\k^\star)^T}
    {s - E^\star(\k)^2} \, \Nc(p^\star) \Bigg] = 0 \, .
    \label{eq:derivation-1}
    \end{align}
The two p.v. integrals can be combined into one and replaced by a sum (up to exponentially suppressed corrections) since the integrated function is regular,
    \begin{align}
    \label{eq:appc1}
    & {\rm p.v.} \! \int\! \frac{\diff \k}{(2\pi)^3} \, \Lc(\k,\P) \, \frac{\Yc^*(\k^\star) \otimes \Yc(\k^\star)^T }{s - E^\star(\k)^2} \, \left[ \Nc(k^{\star}) - \Nc(q^{\star}) \right] \\
    & = \frac{1}{L^3} \sum_{\k} \, \Lc(\k,\P) \, \frac{\Yc^*(\k^\star) \otimes \Yc(\k^\star)^T }{s - E^\star(\k)^2} \, \left[ \Nc(k^{\star}) - \Nc(q^{\star}) \right] + \Oc(e^{-m_\pi L}) \, .
    \label{eq:appc2}
    \end{align}
The term proportional to $\Nc(p^\star)$ cancels with the momentum sum in the second line of Eq.~\eqref{eq:derivation-1}. It leads to,
    \begin{align}
    \det\Bigg[ & \one + 
    \frac{\xi}{L^3}\sum_{\k} \, \Lc(\k,\P) \, \frac{ \Yc^*(\k^\star) \otimes \Yc(\k^\star)^T }{s - E^\star(\k)^2} \, \Nc(k^{\star}) \Bigg] = 0 \, .
    \end{align}
which is the quantization condition proposed in Eq.~\eqref{eq:d_L_dispersive}. This completes the demonstration of its equivalence to the L\"uscher condition for energies above the nearest left-hand singularity.

\subsection{Parametrizations of $\Nc$}
\label{app:n-params}

In the following, we suggest two example models for $\Nc_\ell$ that may become useful in practical applications.

\subsubsection{The pole model}

The simplest non-trivial model for the left-hand singularity structure of a partial-wave amplitude includes a finite number of poles that mimic a branch cut,
    \begin{align}
    \im \Mc_\ell(s) = - \pi \sum_{i=0} g_i \, \delta(s - s_i) \, ,
    \end{align}
where $s_i$ are (ordered) positions of the poles, $s_{i+1} < s_i$, and $-\pi g_i$ are their residues. Models of this type, although simplistic, were shown to approximate well certain types of long-range interactions~\cite{bjorken1960test, ALY1966241, kamal1966s, KOK1972300, KOK1973386} and in some cases can be related to the effective-range expansion of the K matrix~\cite{burkhardt1969dispersion}. Assuming a contribution from just one pole at $s_0 = s_{\rm lhc}$, the solution of the equation for $\Nc_\ell(s)$ depends on a single constant, $g_0 = g$,
    \begin{align}
    \label{eq:pole-model}
    \Nc_{\ell}(s) = \frac{g \, \Dc_\ell(s_0)}{s - s_{\rm lhc}} \, .
    \end{align}
Once the $N/D$ quantization condition, Eq.~\eqref{eq:ND_QC}, fixed its value from the known lattice QCD energy levels, Eq.~\eqref{eq:pole-model} would be then used in the (unsubtracted) equation for $\Dc_\ell(s)$, Eq.~\eqref{eq:D_int_eq}. After ensuring consistency of the solution at $s=s_{\rm lhc}$, it gives, $\Dc_{\ell}(s) = 1 - \lambda I_\ell(s)/(1 + \lambda I_\ell(s_{\rm lhc}))$, where $I_\ell$ is the integral, 
    \begin{align}
    I_\ell(s) = \int_{s_{\mathrm{thr}}}^{\infty} \! \frac{\diff s'}{\pi} \, \frac{\rho_\ell(s') }{(s' - s - i\epsilon) (s' - s_{\rm lhc})} = 
    - \frac{1}{\pi} \, \rho_\ell(s) \, 
    \frac{1}{(s-s_{\rm lhc})} \, \log \left(\frac{s - (m_1+m_2)^2}{s_{\rm lhc} - (m_1+m_2)^2} \right) \, .
    \end{align}
The final amplitude is,
    \begin{align}
    \label{eq:model0amplitude}
    \Mc_\ell^{-1}(s) = \frac{s-s_{\rm lhc}}{g} - \rho_\ell(s) \, \frac{1}{\pi}  \log \left(\frac{s - (m_1+m_2)^2}{s_{\rm lhc} - (m_1+m_2)^2} \right) \, .
    \end{align}
It has the correct right-hand branch cut and a built-in left-hand pole singularity at $s=s_{\rm lhc}$. In addition, it has dynamical poles given by the zeros of Eq.~\eqref{eq:model0amplitude}.

\subsubsection{The one-particle exchange model}

A more sophisticated choice is to consider a $t$-channel, one-particle exchange (OPE) amplitude,
    \begin{align}
    \Tc(s,t) = \frac{g^2}{t - \mu^2 + i\epsilon} \, ,
    \end{align}
where $t = -2q^{\star\,2} (1-\cos\theta^\star)$, $g$ is a coupling, and $\mu$ is the exchanged particle mass~\cite{PhysRev.136.B1120, Du:2024snq}. Using the completeness condition,
    \begin{align}
    \frac{1}{z' - z} = \sum_{\ell = 0}^\infty (2\ell+1) \, Q_\ell(z') P_\ell(z) \, ,
    \end{align}
where $P_\ell$ and $Q_\ell$ are the Legendre functions of the first and second kind, respectively, we find the partial-wave-projected amplitude,
    \begin{align}
    \label{eq:ope_pw}
    \Tc_\ell(s) = -\frac{g^2}{2p^{\star\,2}}\,Q_\ell(\zeta + i\epsilon) \, ,
    \end{align}
with $\zeta = 1 - \mu^2 / 2p^{\star\,2}$. This amplitude is analytic everywhere in the complex $s$ plane, except for a branch cut with branch points defined by $\zeta = \pm 1$. In choosing a parametrization for $\Nc_\ell$, the most rigorous construction is provided by the integral equation for $\Nc_\ell$ itself, such as Eq.~\eqref{eq:proof1}. We note here that if one were to study a coupled channel system, e.g., in the partial waves such as $NN$ or in particle species such as $\pi\pi, K\bar{K}$, then in principle, one must make a model for $\Tc$ and feed it through Eq.~\eqref{eq:proof1} to maintain the symmetry of $\Mc$. For single channel systems, such as those discussed in this work, one may choose instead to use Eq.~\eqref{eq:ope_pw} as motivation for the structural form of $\Nc$. In other words, one can consider a parametrization such as,
    \begin{align}
    \Nc_\ell(s) = \Rc_\ell(s) \, Q_\ell(\zeta+i\epsilon) \, ,
    \end{align}
where $\Rc_\ell$ is a rational function in $s$ (or $q^{\star\,2}$) provided it is chosen to ensure convergence of the dispersion integral at $s\to\infty$, as was required by our derivation. This function absorbs not only any residual short-distance effects associated with couplings of OPE but also local interactions that we do not control. The $Q_\ell$ function inherits the OPE cut structure, which is kinematically known and thus fixed.

Note that near the threshold $p^\star \sim 0$, $\zeta \sim p^{\star\,-2}$, thus $Q_\ell \sim p^{\star\,2\ell+2}$. So, the partial-wave OPE behaves as $\Tc_\ell \sim (p^{\star})^{2\ell}$. In the high-energy limit, $p^\star \sim\sqrt{s}/2$ and $\zeta \sim 1 + O(s^{-1/2})$, thus $Q_\ell\sim \log(s)$, and therefore $\Tc_\ell \sim \log(s)/s$ and one is forced to either include subtractions as discussed in~\ref{app:subtractions}, or parameterize $\Rc_\ell \sim s^{-\ell}$ in this limit. Once the parameters of $\Rc_\ell$ are fixed from the lattice QCD energies via Eq.~\eqref{eq:ND_QC}, then the amplitude is reconstructed from Eqs.~\eqref{eq:ND_decomp} and~\eqref{eq:D_int_eq}.

\end{document}


\preprint{JLAB-THY-24-4216}

\newcommand{\wm}{Department of Physics, William \& Mary, Williamsburg, VA 23187, USA}
\newcommand{\uw}{Department of Physics, University of Washington, WA 98195, USA}
\newcommand{\ceem}{Center for  Exploration  of  Energy  and  Matter,  Indiana  University,  Bloomington,  IN  47403,  USA}
\newcommand{\indiana}{Physics  Department,  Indiana  University,  
Bloomington,  IN  47405,  USA}
\newcommand{\jlab}{Theory Center, Thomas  Jefferson  National  Accelerator  Facility,  Newport  News,  VA  23606,  USA}


\title{Supplementary material for: \\
Finite-volume quantization condition from the $N/D$ representation}


\author{Sebastian~M.~Dawid\orcidlink{0000-0001-8498-5254}}
\email[email: ]{dawids@uw.edu}
\affiliation{\uw}

\author{Andrew~W.~Jackura\orcidlink{0000-0002-3249-5410}}
\email[email: ]{awjackura@wm.edu}
\affiliation{\wm}

\author{Adam~P.~Szczepaniak\orcidlink{0000-0002-4156-5492}}
\email[email: ]{aszczepa@indiana.edu}
\affiliation{\jlab}
\affiliation{\indiana}
\affiliation{\ceem}


\begin{abstract}
This supplement consists of five short sections.
\end{abstract}

\date{\today}
\maketitle

\section{Formal aspects of the finite-volume amplitudes}
\label{app:subtractions}

Some of these can be added to the main text, some shouldn't.

\begin{enumerate}
    \item discuss the formal definition of the finite-volume transition operator and its unitarity (as defined in the footnote),
    \item discuss the p.w. projection of the FV amplitude and truncation of the angular momentum basis,
    \item discuss that the FV amplitude is not purely real above the threshold, it has a structure of the Dirac delta functions,
    \item discuss the convergence of the infinite-volume sum.
\end{enumerate}

The FV transition operator is defined in complete analogy to its IV version. Namely, we define a theory in a box, $H_L = H_{0,L} + V_{L}$, where $H_{0,L}$ is the free and $V_{L}$ is the interaction Hamiltonian. The transition operator is $T_L(z) = V_L + V_L \, G(z) \, V_L$, where $G(z) = (z-H_L)^{-1}$ is the Hamiltonian's resolvent. By construction, the poles of $T_L$ in $z$ coincide with the spectrum of the theory. The FV scattering amplitude, $\Mc_L$, is the matrix element of $T_L$.

The FV scattering amplitude is defined as a matrix element of the evolution operator taken between (discrete) momentum eigenstates of the asymptotically free particles (hadrons) moving inside a large box. Periodic boundary conditions mean that the three-dimensional coordinate space if topologically equivalent to  a torus. The size of the torus is assumed to be much larger then the range of interactions  so that hadrons are  effectively free, but  because of the boundary condition,  momenta to be discrete given by   ${\bf p }  = 2\pi {\bf n}/L$. 
~\andrew{I don't quite understand what this means? There is no notion of asymptotically free in a finite volume. Do you mean that the range of interactions is such that they do not experience severe FV corrections? This means that the only FV effects come from hadron propagation, and we can ignore these other effects (ala self energy of a hadron)}.
\color{red} The statement "There is no notion of asymptotically free in a finite volume" cannot be correct. 1) for example we are using asymptotically free states when  we compute the sum over states on the r.h.s of unitarity relation 2) or when one writes the free propagators in the EFT evaluation of loops 3) finally the FV in coordinate space mans scattering inside a 3dim tours and outside the range of interaction particles  move freely. (There would be no asymptotic state if the 
FV meant hard wall at the box boundary ie. vanishing of the  wave function on the wall, but his is not the case here.) 
\color{black} 
The evolution operator is unitary, which implies the imaginary part of on-shell transition amplitude ${\cal M}$  satisfies Eq.~\ref{eq:IV-unitarity}. In IV, Lorentz symmetry implies that the amplitude is a function of two invariants, which can be conventionally chosen as the magnitude of the relative momenta (each is fixed by the c.m energy) and the angle between them, {\it aka}  the c.m scattering angle.  
The box in the FV breaks Lorentz symmetry and the scattering amplitude depends on the center of mass momentum (measured relative to the box). Furthermore, since the rotational symmetry is broken, the amplitude depends on the angles that define the direction of the relative momenta measured in the box fixed frame.  It is assumed, however, that active Lorentz transformations of the particle states in the box fixed frame remain unbroken~\andrew{...I don't agree with this, Lorentz transformation in a box are broken period. So, what do you mean here?}. 
\color{red} Isn't it true that going from $\bf p$ to ${\bf p}^*$ one uses the some form of Lorentz transformation as in IV  ? \color{black} 
The magnitudes of each relative momentum and of the total momentum 
 are restricted by the periodic boundary condition imposed in individual particle momenta in the rest frame of the box. So finally we have ${\cal M}_L =
  \Mc_{L;\ell'm'_\ell,\ell m_\ell}(p^\star; \P)$. Because the set of relative  momenta is discrete, the partial  waves defined in  Eq.~\ref{eq:FV-PW}, 
 unlike in the IV, the FV partial wave amplitude depends on two sets of spin quantum numbers, $lm$ and $l'm'$.  
We denote the discrete sets of enumerating  the partial wave wave quantum numbers by Greek letters  {\it e.g.}  $\alpha = lm$, $\alpha' = l'm'$. The  discrete 
 set's of spherical angles describing the directions of 
  momenta $\hat{n}^* = \p^*/|\p^*|$ and $\hat{n'}^\star =
  \p'^*/|\p'^*|$   are labeled  by the 
   the vectors on integers $\n$ and $\n'$ corresponding to the relative particle momenta at finite $\P$ ($p'^* = |{\bf p'}^*|$, $p^*
 = |{\bf p}^*|$) ,

    \begin{align}
    \Mc_L(\p',\p; \P) & = 4\pi\, \sum_{\alpha,\beta 
     }  Y_{\alpha }(\hat{n}'^\star) \, 
    \Mc_{L;\alpha,\beta}(p'^*,p^*, \P) \, 
    Y_{\beta}^*(\hat{n}^\star) \, .
    \end{align}
and 
    \begin{align}
     \label{eq:FV-PW-inv}
& 
     \Mc_{L;\alpha,\alpha'}(p'^*,p^\star; \P)
      = 
\frac{1}{4\pi} \sum_{\beta',\beta} C^{-1}_{\alpha',\beta'} 
 \left[\sum_{{\bf n},{\bf n'}} Y_{\beta'}(\hat{n}'^*) 
  \Mc_L(\p',\p; \P) Y^*_{\beta}(\hat n^*)  \right]
C^{-1}_{\beta,\alpha}
 \end{align}
where the matrix $C^{-1}$ is the inverse of matrix $C$ defined by 
\begin{equation} 
 C_{\alpha,\beta} = \sum_{\hat n} Y^*_{\alpha}(\hat n) Y_{\beta}(\hat n)   
\end{equation}
which, in the IV ($\sum_{\n} \to \int d\hat{\n}$)   
is equal to the identify matrix but it is not in the FV. 

\begin{equation} 
\sum_{\gamma}C^{-1}_{\alpha\gamma}C_{\gamma,\beta} = \delta_{\alpha,\beta} 
\end{equation} 
As an example consider the FV partial wave projection of the amplitude representing exchange of a stable particle of mass $\mu$
 in the $t$ channel. This amplitude is proportional to the function 
 \begin{equation} 
 f(\hat n'^*,\hat n^*) = \frac{1}{z(\mu,E) - \hat n'^*\cdot \hat n^*}
 \end{equation} 
 with $z>1$ for $E$ above  threshold. Independent from the fact that in the FV $n'$ and $\n$ take  only discrete values, 
 $f$ can be expanded as a power series in the scalar product yielding 
 \begin{equation} 
  f(\hat n'^*,\hat n^*) = \sum_{lm} (2l+1) Q_l(z) P_l( \hat n'^*\cdot \hat n^* ) = 4\pi \sum_{\gamma}  Q_\gamma(z)  Y^*_{\gamma}(\hat n'^*)  Y_\gamma(\hat n^*) 
  \end{equation} 
  Substituting this for ${\cal M}_L$ in the right hand side of Eq.~\ref{eq:FV-PW-inv}
 one finds for the FV partial wave projection of $f$ 
 \begin{equation} 
 f_{\alpha\beta} = \delta_{\alpha,\beta} Q_\alpha(z)
 \end{equation} 
\color{red} We could comment  that this the same structure as found for example in Max's paper\color{black}
Thus the partial wave amplitude corresponding a particle exchange in the crossed channel is 
 actually diagonal coincides with IV partial wave amplitude.
    It is worth noting that that the above result is consistent with the fact that 
      symmetry braking is entirely kinematical,  and it only affects the phase space. 
    In turn, phase space constrains the amplitude though the unitarity relation. Similarly, the FV affects the phase space of the crossed channel reaction,  as discussed in   Appendix~\ref{app:n-is-nl}. 
The FV partial waves are complex functions of the c.m. energy, as implied by unitraity. In fact when the energy is chosen to be equal the a sum of energies on non-interacting particles the imaginary part becomes infinite. \andrew{I would add that for the FV amplitude, the imaginary part is simply the singular behavior of the poles from the delta functions. One must choose an $i\epsilon$ convention to close contours, and this is why the imaginary part has these delta functions.} 
\color{red} I would only like to stress that $i\epsilon$ is there to begin  with (comes from the definition of the evolution $U(-\infty,\infty)$ \color{black} The same is true in other formulations, {\it e.g.} based, for example on EFT's and using FV Feynman amplitudes. Therefore, the implication of FV unitarity is that, when considered as an analytical function of the c.m energy, as in general, demanded by causality, the FV partial wave amplitudes have poles on the real axis instead of a cut. Since the physical energies of interacting particles in the box 
 always differ from those of the noninteracting system, the partial wave amplitude (and the QC) are real at these energies. 
In addition to the unitarity poles, the FV partial wave amplitude inherits singularities from the crossed channels. 
Since these are also constrained by unitarity, in the FV one thus expects poles instead of cuts when the amplitude is considered as a function of the cross-channel Mandelstam variables. However, the partial wave projection turns these poles into cuts~\andrew{We need to be careful with 'projection', as this is still uncertain to me for a pure FV quantity (the cubic harmonics, which are subductions of the spherical harmonics for FV irreps, have a projection which is sums). I think of it as more of a particular partial wave matrix element has a piece of the pole, which is the log. I know this amounts to the same thing, but I'm trying to not make things too confusing.}.
\color{red} I think the equations above should clarify how the projections  works \color{black}
 In summary, the partial wave FV amplitude, when considered as an analytical function of the energy, has right-hand poles (instead of the right-hand cut), left-hand cuts, and possibly bound state poles on the real axis below the elastic threshold. The dispersion relation for the partial waves and for  the $N$ and $D$ functions are derived based on this singularity structure. 

Note that near the threshold $p^\star \sim 0$, $\zeta \sim p^{\star\,-2}$, thus $Q_\ell \sim p^{\star\,2\ell+2}$. So, the partial-wave OPE behaves as $\Tc_\ell \sim (p^{\star})^{2\ell}$. In the high-energy limit, $p^\star \sim\sqrt{s}/2$ and $\zeta \sim 1 + O(s^{-1/2})$, thus $Q_\ell\sim \log(s)$, and therefore $\Tc_\ell \sim \log(s)/s$ and one is forced to either include subtractions as discussed in~\ref{app:subtractions}, or parameterize $\Rc_\ell \sim s^{-\ell}$ in this limit. Once the parameters of $\Rc_\ell$ are fixed from the lattice QCD energies via Eq.~\eqref{eq:ND_QC}, then the amplitude is reconstructed from Eqs.~\eqref{eq:ND_decomp} and~\eqref{eq:D_int_eq}.

\section{Subtracted $N/D$ equations}
\label{app:subtractions}

To ensure (or improve) convergence of the $N/D$ integrals, it is common to introduce subtractions. One writes,
    \begin{align}
    \label{eq:D_int_eq1}
    \Dc_\ell(s) & = d_\ell^{(j)}(s) - (s-s_0)^j \int_{s_{\mathrm{thr}}}^{\infty} \! \frac{\diff s'}{\pi} \, \frac{\rho_\ell(s')}{(s' - s - i\epsilon) (s' - s_0)^j } \, \Nc_\ell(s') \, , \\
    \label{eq:N_int_eq1}
    \Nc_\ell(s) & = n_{\ell}^{(i)}(s) + (s-s_0)^i \int_{-\infty}^{s_{\rm lhc}}\!\frac{\diff s'}{\pi} \, \frac{\im \Mc_\ell(s')}{(s' - s - i\epsilon) (s' - s_0)^i } \, \Dc_\ell(s') \, ,
    \end{align}
where $i,j \in \mathbb{N}$ and $n_\ell^{(i)}$, $d_\ell^{(j)}$ are some polynomials of the $i$th and $j$th order, respectively. The choice of the subtraction point $s_0$ redefines $\Nc$ and $\Dc$ by a constant multiplicative factor. In many cases of practical interest, the simplest, single subtraction in the dispersive integral for $\Dc_\ell$ is enough to restore convergence of the $N/D$ representation. Focusing on such an example, we write,
    \begin{align}
    \Dc_\ell(s) & = 1 - \int_{s_{\mathrm{thr}}}^{\infty} \! \frac{\diff s'}{\pi} \, \frac{s}{s'} \, \frac{\rho_\ell(s')}{(s' - s - i\epsilon)} \, \Nc_\ell(s') \, , \\
    \label{eq:N_int_eq2}
    \Nc_\ell(s) & = \int_{-\infty}^{s_{\rm lhc}}\!\frac{\diff s'}{\pi} \, \frac{\im \Mc_\ell(s')}{s' - s - i\epsilon } \, \Dc_\ell(s') \, ,
    \end{align}
where we picked $s_0 = 0$. Following the logic of the main text, it can be shown that the corresponding quantization condition is given by Eq.~\eqref{eq:ND_QC} with,
    \begin{align}
    \Dc_{L}(p^\star,\P) = \one - \frac{\xi}{L^3} \sum_{\k} 
    \frac{s \, \Lc(\k,\P) }{E^\star(\k)^2} \, 
    \frac{\Yc^*(\hat{\k}^\star)
    \otimes 
     \Yc(\hat{\k}^\star)^T }
    {s - E^\star(\k)^2} \,
    \Nc(k^\star) \, .
    \label{eq:d_L_dispersive1}
    \end{align}
Note that the additional factor does not affect the proof that $\Nc = \Nc_L$ for physical kinematics presented in~\ref{app:n-is-nl}, nor the equivalence of our formalism with the L\"uscher quantization condition derived in~\ref{app:Luscher}. This is because the factor $s/E^\star(\k)^2$ can be replaced by $1$ under every sum-integral difference appearing in these appendices.

\section{Proof that $\Nc = \Nc_L + \Oc(\exp(- \Delta L))$}
\label{app:n-is-nl}

\begin{figure}[t!]
    \centering
    \includegraphics[width=0.8\textwidth]{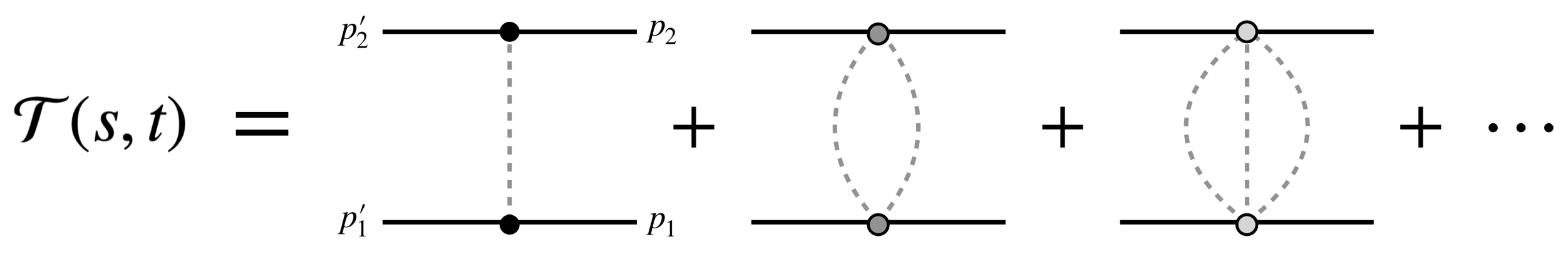}
    \caption{Example Feynman diagrams contributing to the left-hand singularity structure of the partial-wave amplitude in some generic effective field theory. Since particles in loops can not go on-shell for the physical energies, these diagrams do not lead to the right-hand singularities.}
    \label{fig:diags}
\end{figure}

Let us first consider the IV partial-wave amplitude $\Mc = \Nc \Dc^{-1}$, where each object is a diagonal matrix in the $(\ell' m_\ell'; \ell m_\ell)$ space. For simplicity, we assume that the left-hand singularities of the amplitude $\Mc$ are given by one- and multi-particle exchange diagrams as shown in Fig.~\ref{fig:diags}. It is not necessary for the derivation but allows us to focus on an example of physical interest. Contributions for such terms are denoted collectively by $\Tc(s,t)$. The $\ell$-th partial-wave of $\Tc(s,t)$ in the $s$ channel is called $\Tc_{\ell}(s)$ and is a matrix element of the (diagonal) matrix $\Tc(s)$. Since $\im \Mc(s) = \im \Tc(s)$ for $s < s_{\rm lhc}$ and  $\im \Tc(s) = 0$ for $s>s_{\rm thr}$, one may write,
    \begin{align}
    \Nc(s) & = \int_{-\infty}^{s_{\rm lhc}} \frac{ds'}{\pi} \frac{\im \Tc(s')}{s' - s - i \epsilon} 
    - \int_{-\infty}^{s_{\rm lhc}} \frac{ds'}{\pi} \frac{\im \Tc(s') }{s' - s - i \epsilon} 
    \int_{s_{\rm thr}}^{\infty} \frac{ds''}{\pi} \frac{\rho_\ell(s'') \, \Nc(s'') }{s'' - s' - i \epsilon} \\
    & = \Tc(s) 
    + \int_{s_{\rm thr}}^{\infty} \frac{ds''}{\pi} \frac{\rho_\ell(s'') \, \im \big[ \Tc(s'') -  \Tc(s) \big] }{s'' - s - i \epsilon} \, \Nc(s'') \, , 
    \label{eq:proof1}
    \end{align}
where we have used the integral equation for $\Dc_\ell$ in the formula for $\Nc_\ell$, and integrated over $s'$ in the second line. As a result, we obtain a single integral equation for $\Nc_\ell(s)$, which can be solved once $\Tc$ is specified. In the following, we assume that all integrals converge. Rigorously, it is incorrect for the type of $\Tc$ contributions considered here (which lead to logarithmic divergences in the $N/D$ equations.) Divergences are taken care of by introducing subtractions, as discussed in~\ref{app:subtractions}, and we proceed without them to maintain the simplicity of the expressions in this argument.

In the FV theory, the $\Nc_L$ is obtained from dispersing its imaginary part $\im \Nc_L = \im \Mc_L \, \Dc_L$. Thus, taking $\Dc_L$ given by Eq.~\eqref{eq:d_L_dispersive-1} and neglecting exponentially suppressed corrections, we obtain,
    \begin{align}
    \Nc_L(p^\star,\P) = \Tc(s) 
    + \sum_{\k} \, \Lc(\k, \P) \, \frac{ \left[\Yc^*(\k^\star) \otimes \Yc(\k^\star)^T \right] \im 
    \big[ \Tc(E^\star(\k)^2) -  \Tc(s) \big]}{E^\star(\k)^2 - s - i \epsilon} \, \Nc_L(k^\star, \P) \, .
    \label{eq:proof2}
    \end{align}
As usual, $p^\star$ is the momentum corresponding to the total invariant mass $s$. To obtain the above expression we performed similar operations as in Eq.~\eqref{eq:proof1}, assumed $p^\star > 0$ (physical kinematics,) and used the fact that $\Tc_L(p^\star) = \Tc(p^\star) + \Oc(e^{-\Delta L})$, where $\Delta$ represents the lowest energy scale in the problem (e.g., pion mass.) 

Note that neither $\Nc_L$ nor $\Tc$ have singularities for physical momenta over which one performs the sum in Eq.~\eqref{eq:proof2}. Moreover, the difference of two $\Tc$'s in the numerator regulates the pole singularity from the vanishing energy denominator. Therefore, by the Poisson summation formula, the sum may be replaced by the corresponding IV integral over momenta---up to the corrections exponentially suppressed with the box size. A similar argument is used to show that~\eqref{eq:appc1} is equal to~\eqref{eq:appc2} in~\ref{app:Luscher}. By variable transformation, one rewrites it as an integral over $s'$, identical to the one appearing in Eq.~\eqref{eq:proof1}. It shows that $\Nc = \Nc_L + \Oc(e^{-\Delta L})$ in the physical energy region, which was necessary for the derivation of the $N/D$ quantization condition in the main text, Eq.~\eqref{eq:d_L_dispersive}. 

\sebastian{Add example with how the poles become a logarithm.}

\section{Reduction to the L\"uscher formalism} 
\label{app:Luscher}

The L\"uscher quantization condition can be expressed as,
    \begin{align}
    \det_{\ell m_\ell} \Big[\Kc^{-1}(s) + F(P,L)\Big] = 0 \, ,
    \label{eq:luscher}
    \end{align}
which holds for the FV spectrum for a given momentum $\P$ and box size $L$. Here, $\Kc$ is the IV two-body $K$ matrix, $[\Kc]_{\ell' m_\ell'; \ell m_\ell} = \delta_{\ell' \ell} \, \delta_{m_\ell' m_\ell} \, \Kc_{\ell}(s)$, related to the scattering amplitude via,
    \begin{align}
    \Mc_{\ell}^{-1}(s) = \Kc_\ell^{-1}(s) - i \rho(s) \, ,
    \end{align}
where $\rho$ is the phase space factor without barrier factors, $\rho(s) = \xi p^\star / 8\pi \sqrt{s}$.
Finally, $F(P,L)$ is a matrix of elements~\cite{Briceno:2015tza},
    \begin{align}
    F_{\ell' m_\ell' ; \ell m_\ell}(P,L) = \xi \left[\frac{1}{L^3}\sum_{\k} - {\rm p.v.} \int\! \frac{\diff^3\k}{(2\pi)^3}\right] \,
    \left(\frac{k^\star}{p^\star}\right)^{\ell'} 
    \frac{(4\pi) \, 
    Y^*_{\ell'm_{\ell'}}({\hat{\k}}^\star) \,
    Y_{\ell m_{\ell}}(\hat{\k}^\star)}
    {2\omega_1(\k) \, 2 \omega_2(\P-\k) \,  \big( E - \omega_1(\k) - \omega_2(\P-\k) \big)} 
    \left(\frac{k^\star}{p^\star}\right)^{\ell} \, ,
    \end{align}
where ``p.v.'' indicates the Cauchy principal value prescription. 

To compare our and L\"uscher's quantization condition, several tedious manipulations are necessary. First, we note that the $N/D$ representation introduced in the main text relates the spectrum to the amplitude $\tilde{M}$, stripped off the threshold barrier factors,
    \begin{align}
    \tilde{\Mc}_{\ell}^{-1}(s) = \tilde{\Kc}_\ell^{-1}(s) - i \rho_\ell(s) \, ,
    \end{align}
where $\tilde{\Kc}_\ell = (p^\star)^{2\ell} \,  \Kc_\ell$. Removing these factors from the quantization condition gives, 
    \begin{align}
    \det_{\ell m_\ell} \Big[\tilde{\Kc}^{-1}(s) + \tilde{F}(P,L)\Big] = 0 \, ,
    \label{eq:luscher1}
    \end{align}
where,
    \begin{align}
    \tilde{F}_{\ell' m_\ell' ; \ell m_\ell}(P,L) = \xi \left[\frac{1}{L^3}\sum_{\k} - {\rm p.v.} \int\! \frac{\diff^3\k}{(2\pi)^3}\right] \, 
    \frac{(4\pi) \, 
    (k^\star)^{\ell'} \, 
    Y^*_{\ell'm_{\ell'}}({\hat{\k}}^\star) \,
    Y_{\ell m_{\ell}}(\hat{\k}^\star) \, 
    ( k^\star)^{\ell} }
    {2\omega_1(\k) \, 2 \omega_2(\P-\k) \, 
    \big(E - \omega_1(\k) - \omega_2(\P-\k) \big)} \, .
    \end{align}
In the next step, we modify the energy denominator of the above function. We add and subtract $(E + \omega_1(\k) + \omega_2(\P-\k))^{-1}$ to write,
    \begin{align} 
    \frac{1}
    {E - \omega_1(\k) - \omega_2(\P-\k)} = \frac{2E}
    {E^2 - (\omega_1(\k) + \omega_2(\P-\k))^2  } - \frac{1}
    {E + \omega_1(\k) + \omega_2(\P-\k)} \, ,
    \end{align}
under the sum-integral difference. The contribution from the second term is exponentially suppressed with the box size, $L$, as it does not contain singularities in the elastic region. We drop it and use the c.m.~variables in the numerator, replacing, $E^2 - (\omega_1(\k) + \omega_2(\P-\k))^2 = \gamma^2 \big( s -  (\omega_1(\k^\star) + \omega_2(\k^\star))^2 \big)$, where $\gamma = E/E^\star$,
    \begin{align}
    \frac{2E}
    {E^2 - (\omega_1(\k) + \omega_2(\P-\k))^2  } = 2 \, \frac{(E^\star)^2}{E} \, \frac{1}
    {s - (\omega_1(\k^\star) + \omega_2(\k^\star))^2  } \, .
    \end{align}
Finally, we add and subtract $(\omega_1(\k^\star) + \omega_2(\k^\star))^2/\sqrt{(\omega_1(\k^\star) + \omega_2(\k^\star))^2 - \P^2}$ from the first fraction. Again, this results in an expression in which some contributions cancel the pole in the second fraction and, thus, are exponentially suppressed with volume under the sum-integral difference. Neglecting them, one can write,
    \begin{align}
    \tilde{F}(P,L) = \xi \left[\frac{1}{L^3}\sum_{\k} - {\rm p.v.} \! \int\! \frac{d\k}{(2\pi)^3}\right] \,
    \Lc(\k,\P) \,
    \frac{ 
    \Yc^*({\k}^\star) 
    \otimes
    \Yc(\k^\star)^T }
    {s - E^\star(\k)^2}  \, ,
    \label{eq:tilde_F}
    \end{align}
where we used the notation introduced in the main text. 

We now compare the IV $K$ matrix and $N/D$ parametrization for $s > s_{\rm thr}$. We start from,
    \begin{equation}
    \tilde{\Kc}^{-1}(s) - i \, Q(p^\star) \, \rho(s) \, Q(p^\star) = \Dc(s) \, \Nc^{-1}(s) \, .
    \end{equation}
where we consider all the IV objects as diagonal matrices in the $(\ell' m_\ell; \ell m_\ell)$ space. The $K$ matrix is given by the real part of the integral representation of $\Dc$, Eq.~\eqref{eq:D_int_eq},
    \begin{equation}
    \tilde{\Kc}^{-1}(s) = \left[ \one - {\rm p.v.} \! \int \frac{ds'}{\pi} \, \frac{Q(k^\star) \rho(s') \, Q(k^\star) \, \Nc(s') }{s' - s} \right] \, \Nc^{-1}(s) \, ,
    \end{equation}
where $k^\star$ is a c.m.~momentum corresponding to the integration variable $s'$. We change the integration variables to momenta, adding a redundant angular part,
    \begin{align}
    \tilde{\Kc}^{-1}(p^\star) = \left[ \one - \xi \, {\rm p.v.} \! \int \frac{d\k^\star}{(2\pi)^3} \, \Lc(\k,\P) \, \frac{ \Yc^*(\k^\star) \otimes \Yc(\k^\star)^T \Nc(k^\star) }{E^\star(\k)^2 - s} \right] \, \Nc^{-1}(p^{\star}) \, .
    \end{align}
Here, we expressed the $\Kc$ and $\Nc$ as functions of momenta. We use the above representation and Eq.~\eqref{eq:tilde_F} in the L\"uscher quantization condition. Multiplying by $\det \Nc(p^\star)$ yields,
    \begin{align}
    \nonumber
    \det_{\ell m_\ell} \Bigg[ & \one + \xi \, {\rm p.v.} \! \int \frac{d\k}{(2\pi)^3} \, \Lc(\k,\P) \, \frac{\Yc^*(\k^\star) \otimes \Yc(\k^\star)^T }{s - E^\star(\k)^2} \, \Nc(k^\star) \\
     & + 
     \xi \left[\frac{1}{L^3}\sum_{\k} - {\rm p.v.} \! \int\! \frac{d\k}{(2\pi)^3}\right] \,
    \Lc(\k,\P) \,
    \frac{\Yc^*(\k^\star) \otimes \Yc(\k^\star)^T}
    {s - E^\star(\k)^2} \, \Nc(p^\star) \Bigg] = 0 \, .
    \label{eq:derivation-1}
    \end{align}
The two p.v. integrals can be combined into one and replaced by a sum (up to exponentially suppressed corrections) since the integrated function is regular,
    \begin{align}
    \label{eq:appc1}
    & {\rm p.v.} \! \int\! \frac{\diff^3\k}{(2\pi)^3} \, \Lc(\k,\P) \, \frac{\Yc^*(\k^\star) \otimes \Yc(\k^\star)^T }{s - E^\star(\k)^2} \, \left[ \Nc(k^{\star}) - \Nc(q^{\star}) \right] \\
    & = \frac{1}{L^3} \sum_{\k} \, \Lc(\k,\P) \, \frac{\Yc^*(\k^\star) \otimes \Yc(\k^\star)^T }{s - E^\star(\k)^2} \, \left[ \Nc(k^{\star}) - \Nc(q^{\star}) \right] + \Oc(e^{-m_\pi L}) \, .
    \label{eq:appc2}
    \end{align}
The term proportional to $\Nc(p^\star)$ cancels with the momentum sum in the second line of Eq.~\eqref{eq:derivation-1}. It leads to,
    \begin{align}
    \det\Bigg[ & \one + 
    \frac{\xi}{L^3}\sum_{\k} \, \Lc(\k,\P) \, \frac{ \Yc^*(\k^\star) \otimes \Yc(\k^\star)^T }{s - E^\star(\k)^2} \, \Nc(k^{\star}) \Bigg] = 0 \, .
    \end{align}
which is the quantization condition proposed in Eq.~\eqref{eq:d_L_dispersive}. This completes the demonstration of its equivalence to the L\"uscher condition for energies above the nearest left-hand singularity.

\section{Parametrizations of $\Nc$}
\label{app:n-params}

In the following, we suggest two example models for $\Nc_\ell$ that may become useful in practical applications.

\subsection{The pole model}

The simplest non-trivial model for the left-hand singularity structure of a partial-wave amplitude includes a finite number of poles that mimic a branch cut,
    \begin{align}
    \im \Mc_\ell(s) = - \pi \sum_{i=0} g_i \, \delta(s - s_i) \, ,
    \end{align}
where $s_i$ are (ordered) positions of the poles, $s_{i+1} < s_i$, and $-\pi g_i$ are their residues. Models of this type, although simplistic, were shown to approximate well certain types of long-range interactions~\cite{bjorken1960test, ALY1966241, kamal1966s, KOK1972300, KOK1973386} and in some cases can be related to the effective-range expansion of the K matrix~\cite{burkhardt1969dispersion}. Assuming a contribution from just one pole at $s_0 = s_{\rm lhc}$, the solution of the equation for $\Nc_\ell(s)$ depends on a single constant, $g_0 = g$,
    \begin{align}
    \label{eq:pole-model}
    \Nc_{\ell}(s) = \frac{g \, \Dc_\ell(s_0)}{s - s_{\rm lhc}} \, .
    \end{align}
Once the $N/D$ quantization condition, Eq.~\eqref{eq:ND_QC}, fixed its value from the known lattice QCD energy levels, Eq.~\eqref{eq:pole-model} would be then used in the (unsubtracted) equation for $\Dc_\ell(s)$, Eq.~\eqref{eq:D_int_eq}. After ensuring consistency of the solution at $s=s_{\rm lhc}$, it gives, $\Dc_{\ell}(s) = 1 - \lambda I_\ell(s)/(1 + \lambda I_\ell(s_{\rm lhc}))$, where $I_\ell$ is the integral, 
    \begin{align}
    I_\ell(s) = \int_{s_{\mathrm{thr}}}^{\infty} \! \frac{\diff s'}{\pi} \, \frac{\rho_\ell(s') }{(s' - s - i\epsilon) (s' - s_{\rm lhc})} = 
    - \frac{1}{\pi} \, \rho_\ell(s) \, 
    \frac{1}{(s-s_{\rm lhc})} \, \log \left(\frac{s - (m_1+m_2)^2}{s_{\rm lhc} - (m_1+m_2)^2} \right) \, .
    \end{align}
The final amplitude is,
    \begin{align}
    \label{eq:model0amplitude}
    \Mc_\ell^{-1}(s) = \frac{s-s_{\rm lhc}}{g} - \rho_\ell(s) \, \frac{1}{\pi}  \log \left(\frac{s - (m_1+m_2)^2}{s_{\rm lhc} - (m_1+m_2)^2} \right) \, .
    \end{align}
It has the correct right-hand branch cut and a built-in left-hand pole singularity at $s=s_{\rm lhc}$. In addition, it has dynamical poles given by the zeros of Eq.~\eqref{eq:model0amplitude}.

\subsection{The one-particle exchange model}

A more sophisticated choice is to consider a $t$-channel, one-particle exchange (OPE) amplitude,
    \begin{align}
    \Tc(s,t) = \frac{g^2}{t - \mu^2 + i\epsilon} \, ,
    \end{align}
where $t = -2q^{\star\,2} (1-\cos\theta^\star)$, $g$ is a coupling, and $\mu$ is the exchanged particle mass~\cite{PhysRev.136.B1120, Du:2024snq}. Using the completeness condition,
    \begin{align}
    \frac{1}{z' - z} = \sum_{\ell = 0}^\infty (2\ell+1) \, Q_\ell(z') P_\ell(z) \, ,
    \end{align}
where $P_\ell$ and $Q_\ell$ are the Legendre functions of the first and second kind, respectively, we find the partial-wave-projected amplitude,
    \begin{align}
    \label{eq:ope_pw}
    \Tc_\ell(s) = -\frac{g^2}{2p^{\star\,2}}\,Q_\ell(\zeta + i\epsilon) \, ,
    \end{align}
with $\zeta = 1 - \mu^2 / 2p^{\star\,2}$. This amplitude is analytic everywhere in the complex $s$ plane, except for a branch cut with branch points defined by $\zeta = \pm 1$. In choosing a parametrization for $\Nc_\ell$, the most rigorous construction is provided by the integral equation for $\Nc_\ell$ itself, such as Eq.~\eqref{eq:proof1}. We note here that if one were to study a coupled channel system, e.g., in the partial waves such as $NN$ or in particle species such as $\pi\pi, K\bar{K}$, then in principle, one must make a model for $\Tc$ and feed it through Eq.~\eqref{eq:proof1} to maintain the symmetry of $\Mc$. For single channel systems, such as those discussed in this work, one may choose instead to use Eq.~\eqref{eq:ope_pw} as motivation for the structural form of $\Nc$. In other words, one can consider a function such as,
    \begin{align}
    \Nc_\ell(s) = \Rc_\ell(s) \, Q_\ell(\zeta+i\epsilon) \, ,
    \end{align}
where $\Rc_\ell$ is a rational function in $s$ (or $q^{\star\,2}$) provided it is chosen to ensure convergence of the dispersion integral at $s\to\infty$, as was required by our derivation. This function absorbs not only any residual short-distance effects associated with couplings of OPE but also local interactions that we do not control. The $Q_\ell$ function inherits the OPE cut structure, which is kinematically known and thus fixed.

Note that near the threshold $p^\star \sim 0$, $\zeta \sim p^{\star\,-2}$, thus $Q_\ell \sim p^{\star\,2\ell+2}$. So, the partial-wave OPE behaves as $\Tc_\ell \sim (p^{\star})^{2\ell}$. In the high-energy limit, $p^\star \sim\sqrt{s}/2$ and $\zeta \sim 1 + O(s^{-1/2})$, thus $Q_\ell\sim \log(s)$, and therefore $\Tc_\ell \sim \log(s)/s$ and one is forced to either include subtractions as discussed in~\ref{app:subtractions}, or parameterize $\Rc_\ell \sim s^{-\ell}$ in this limit. Once the parameters of $\Rc_\ell$ are fixed from the lattice QCD energies via Eq.~\eqref{eq:ND_QC}, then the amplitude is reconstructed from Eqs.~\eqref{eq:ND_decomp} and~\eqref{eq:D_int_eq}.

\bibliographystyle{apsrev4-1}
\bibliography{main}